\documentclass[aps,pra,twocolumn,superscriptaddress,showpacs]{revtex4-1}
\usepackage{amsmath,amssymb,color}
\usepackage{graphicx,bm,dcolumn}

\newcommand{\bk}{\mathbf{k}}
\newcommand{\bq}{\mathbf{q}}
\newcommand{\dSe}{\delta S_e}
\newcommand{\half}{{\textstyle\frac{1}{2}}}
\newcommand{\hloc}{h_{\loc}}
\newcommand{\hostimp}{\text{host-imp}}
\newcommand{\host}{\mathrm{host}}
\newcommand{\hrho}{\hat{\rho}}
\newcommand{\imp}{\mathrm{imp}}
\newcommand{\loc}{\mathrm{loc}}
\newcommand{\Mloc}{M_{\loc}}
\newcommand{\pdag}{\phantom{\dag}}
\newcommand{\Qloc}{Q_{\loc}}
\newcommand{\rhoimp}{\hrho_{\imp}}
\newcommand{\Simp}{S_{\imp}}
\newcommand{\Simpz}{\Simp^z}

\begin{document}
\title{Entanglement Entropy Near Kondo-Destruction Quantum Critical Points}
\author{J.\ H.\ Pixley}
\affiliation{Condensed Matter Theory Center and the Joint Quantum Institute,
Department of Physics, University of Maryland,
College Park, Maryland 20742-4111, USA}
\affiliation{Department of Physics and Astronomy, Rice University,
Houston, Texas, 77005, USA}
\author{Tathagata Chowdhury}
\affiliation{Department of Physics, University of Florida, Gainesville, Florida
32611-8440, USA}
\author{M.\ T.\ Miecnikowski}
\altaffiliation[Present address: ]{Department of Physics, University of
Colorado, Libby Drive, Boulder, Colorado 80302-0390, USA}
\affiliation{Department of Physics, University of Florida, Gainesville, Florida
32611-8440, USA}
\affiliation{Department of Applied Physics and Applied Mathematics, Columbia
University, 500 W.\ 120th St., Mudd 200, MC 4701 New York, NY 10027}
\author{Jaimie Stephens}
\affiliation{Department of Physics, University of Florida, Gainesville, Florida
32611-8440, USA}
\affiliation{Department of Physical and Environmental Sciences, Colorado Mesa
University, Grand Junction, Colorado 81501-3122, USA}
\author{Christopher Wagner}
\affiliation{Department of Physics, University of Florida, Gainesville, Florida
32611-8440, USA}
\author{Kevin Ingersent}
\affiliation{Department of Physics, University of Florida, Gainesville, Florida
32611-8440, USA}

\date{\today}
\begin{abstract}
We study the impurity entanglement entropy $S_e$ in quantum impurity models
that feature a Kondo-destruction quantum critical point (QCP) arising from a
pseudogap in the conduction-band density of states or from coupling to a
bosonic bath. On the local-moment (Kondo-destroyed) side of the QCP, the
entanglement entropy contains a critical component that can be related to the
order parameter characterizing the quantum phase transition. In Kondo models
describing a spin-$\Simp$, $S_e$ assumes its maximal value of $\ln(2\Simp+1)$
at the QCP and throughout the Kondo phase, independent of features such as
particle-hole symmetry and under- or over-screening. In Anderson models, $S_e$
is nonuniversal at the QCP, and at particle-hole symmetry, rises monotonically
on passage from the local-moment phase to the Kondo phase; breaking this
symmetry can lead to a cusp peak in $S_e$ due to a divergent charge
susceptibility at the QCP. Implications of these results for quantum critical
systems and quantum dots are discussed.
\end{abstract}
% PACS:
% 71.10.Hf Non-Fermi-liquid ground states, electron phase diagrams and
% phase transitions in model systems
% 71.27.+a Strongly correlated electron systems; heavy fermions
% 75.20.Hr Local moment in compounds and alloys; Kondo effect, valence
% fluctuations, heavy fermions (see also 72.15.Qm Scattering
% mechanisms and Kondo effect)
% 05.10.Cc Renormalization group methods

\maketitle

\section{Introduction}
\label{sec:intro}

Entanglement entropy has emerged as a powerful tool for quantifying correlations
in pure and mixed states of quantum many-body systems, and particularly for
characterizing unconventional ground states. In cases where a local order
parameter is not sufficient, the entanglement entropy $S_e$ [defined in Eq.\
\eqref{S_e:def} below] can be used to identify nontrivial topological order
\cite{topological}. It has been established\cite{JPA.09} that $S_e$ exhibits
nontrivial scaling in the vicinity of continuous zero-temperature
phase transitions, i.e., at quantum critical points \cite{Sachdev}. From an
information perspective, entanglement is a generic feature of quantum-mechanical
systems that may be used for storing information as well as for quantum
computation \cite{Nielsen}.

Due to the nonlocality of entanglement entropy, there are relatively few
interacting problems where it has proved possible to accurately determine $S_e$,
particularly in the vicinity of quantum phase transitions. In this context,
quantum impurity problems have the advantage that they can be solved reliably
via a number of well-established techniques. Conformal field
theory \cite{Affleck.09}, the density-matrix renormalization
group \cite{Affleck.09}, and the numerical renormalization
group \cite{Kopp.07} have all been used to calculate entanglement entropy in
the Kondo and spin-boson models.
Quantum impurity models therefore provide a natural setting for systematic study
of the interplay between quantum criticality and entanglement entropy. For
example, it has been established using the numerical
renormalization-group (NRG) method \cite{Kopp.07,LeHur.07,LeHur.08} that
$S_e$ exhibits a cusp peak at the quantum critical point (QCP) that separates
the localized and delocalized phases of the sub-ohmic spin-boson model.
Whether or not a cusp peak in $S_e$ at the QCP is a generic feature of
continuous impurity quantum phase transitions is an open and intriguing
question.

The Kondo and Anderson impurity models \cite{Hewson.93}, originally shown
to describe the many-body screening of the magnetic moments of dilute magnetic
impurities in nonmagnetic metals, have now found applications in other contexts,
including but not limited to the physics of heavy $f$-electron systems and
transport through quantum dots. Here, we study variants of these models that
exhibit critical destruction of the Kondo effect, where Kondo screening is
suppressed at a second-order quantum phase arising from
the presence of a pseudogap in the
conduction-band density of states around the Fermi energy \cite{Withoff.90,
Bulla.97,Buxton.98,Ingersent.02,Fritz.04,Schneider.11,Glossop.11, Pixley.12,
Zamani.13} and/or an additional coupling of the impurity to a bosonic
environment \cite{Sengupta.00,Zhu.02,Zarand.02,Kircan.04,Glossop.05,Glossop.07,
Pixley.13}. Kondo destruction QCPs in the Bose-Fermi Kondo model have
been proposed to describe anomalous quantum criticality in heavy fermion
metals \cite{Si.01,Si.03} and non-equilibrium criticality in a quantum dot with
ferromagnetic leads \cite{Kirchner.05}. Pseudogap Kondo and Anderson models
have been used to study nonmagnetic impurities in $d$-wave superconductors
\cite{Vojta.01} and in graphene \cite{Fritz.13}. It has also been shown that
tuning a double-quantum-dot system can produce a pseudogap in the effective
density of states \cite{Silva.06,Silva.08}. While each of these models have
been studied in the past, little is known about their entanglement entropy
properties, a gap that this work aims to fill.

By combining analytic and NRG calculations, we establish that the entanglement
entropy between a magnetic impurity and its environment contains a critical
component in the vicinity of these Kondo-destruction QCPs. In Kondo models with
a spin-$\Simp$ impurity moment, we show that $S_e$ takes its maximal value of
$\ln(2\Simp+1)$ at the QCP and throughout the Kondo phase, and decreases in a
power-law fashion on entry into the Kondo-destroyed or local-moment phase. These
behaviors highlight some differences between the Kondo and spin-boson models.
In pseudogap Anderson models, we show that charge fluctuations lead $S_e$
to take a nonuniversal value at the quantum phase transition. At particle-hole
symmetry, $S_e$ always increases with increasing impurity-band hybridization
$\Gamma$ and is therefore never peaked at the the quantum phase transition. Away
from particle-hole symmetry, $S_e$ can vary non-monotonically with $\Gamma$,
and in some cases exhibits a cusp peak at the QCP.

The remainder of this paper is organized as follows:
Section \ref{sec:general} reviews general characteristics of the entanglement
entropy and summarizes the universal behaviors that we find at Kondo-destruction
QCPs. Detailed analysis of various Kondo models and Anderson models is presented
in Secs.\ \ref{sec:Kondo} and \ref{sec:Ander}, respectively. We discuss our
results in Sec.\ \ref{sec:discuss} and conclude in Sec.\ \ref{sec:conclusions}.

\section{General Considerations}
\label{sec:general}

Entanglement entropy captures the degree of quantum nonlocality in the
ground-state wave function. Specifically, it is a property associated with a
partition of the system into two regions $A$ and $B$ that effectively ``cuts''
the ground state along the boundary between the regions.
Upon tracing the system's density operator $\hrho$ over region
$B$, one obtains the reduced density operator in region $A$,
$\hrho_A=\mathrm{Tr}_B \, \hrho$. Similarly, one can trace over
region $A$ to obtain $\hrho_B=\mathrm{Tr}_A \, \hrho$. The
entanglement entropy is the von Neumann entropy of $\hrho_A$ or
$\hrho_B$, i.e.,
\begin{equation}
\label{S_e:def}
S_e(A|B) = -\mathrm{Tr}_A  (\hrho_A \ln \hrho_A)
    \equiv -\mathrm{Tr}_B (\hrho_B \ln \hrho_B) ,
\end{equation}
which measures the extent to which region $A$ is entangled with region $B$.

\begin{figure}[t]
\includegraphics[height=2in]{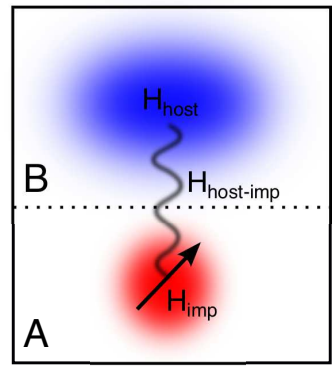}
\caption{\label{fig:schematic}(Color online)
Schematic representation of the model Hamiltonians considered in this
work.  The impurity degrees of freedom $H_{\imp}$
are coupled via $H_{\hostimp}$ (wavy line) to the bath
degrees of freedom $H_{\host}$.
We trace out region B, and determine the entanglement entropy of region A.}
\end{figure}

In quantum impurity problems, the entanglement entropy between the impurity and
the rest of the system is defined by taking region A to contain solely the
impurity degrees of freedom, while region B describes the host (i.e., the rest
of the system), as shown schematically in Fig.\ \ref{fig:schematic}.
Upon tracing out the host, we obtain the impurity reduced
density operator $\rhoimp$ acting in a vector space of dimension $d_{\imp}$.
Equation \eqref{S_e:def} then gives the impurity entanglement entropy
\cite{Kopp.07,LeHur.07}
\begin{equation}
\label{S_e:imp}
S_e=-\sum_{i=1}^{d_{\imp}} p_i \ln p_i,
\end{equation} 
where $\{ p_i \}$ is the set of eigenvalues of $\rhoimp$.
The corresponding eigenstates $\{ |i\rangle \}$ must respect the system's
symmetries, a constraint that allows the eigenvalues $p_i$ to be expressed
in terms of expectation values of impurity operators that can readily be
calculated using the NRG. Since the host degrees of freedom have been
completely traced out, the impurity entanglement entropy measures only the
entanglement between the impurity and the host as a whole. Details of the
host---such as the number, dispersion, and any internal interactions of the
conduction bands and/or the bosonic baths---influence $S_e$ only insofar as
they affect the impurity matrix elements that determine the eigenvalues of
$\rhoimp$.

For a ground state of product form
$|\psi\rangle=|\phi\rangle_{\imp} \otimes |\chi\rangle_{\host}$, one can choose
$p_1 = 1$ and  $p_i = 0$ for all $i > 1$, implying that $S_e=0$. At the other
extreme, a state of maximal entanglement between the impurity and its host
is described by $p_i = 1/d_{\imp}$ for all $i$, leading to $S_e=\ln d_{\imp}$.

A complication arises if the system is not in a pure state, as is likely to be
the case when there is ground-state degeneracy. For example, in the trivial
limit where the impurity and the host are decoupled, $n$-fold degeneracy of the
impurity ground state results in $\rhoimp$ having $n$ values $p_i=1/n$ and
$d_{\imp} - n$ values $p_i = 0$, implying that $S_e = \ln n$. In order to avoid
such misleading indications of entanglement, it is necessary to break the
ground-state degeneracy of the impurity to obtain a pure state.

In the present work, where we treat magnetic impurities, the ground-state
degeneracy can be lifted by the application of an infinitesimal local magnetic
field $\hloc$ that couples solely to the impurity through a Hamiltonian term
$\hloc \Simpz$, where $\mathbf{S}_{\imp}$ is the impurity spin operator and the
Land\'{e} $g$ factor and the Bohr magneton have both been set to unity. For
this reason, we consider a two-parameter function $S_e(x,\hloc)$, where $x$ is
a nonthermal, nonmagnetic parameter that tunes the system through a QCP at
$x=x_c$. In many cases, we employ a reduced variable $\Delta=(x-x_c)/x_c$ such
that the QCP is located at $\Delta=0$. It also proves convenient to define the
local-field-dependent part of the entanglement entropy (note the sign)
\begin{equation}
\dSe(x,\hloc) = S_e(x, 0) - S_e(x, \hloc),
\end{equation}
and to introduce the shorthand notations
\begin{subequations}
\begin{align}
\label{S_e^+:def}
S_e^+(x) &= S_e(x, \hloc=0^+), \\
\label{dS_e:def}
\dSe(x) &= \dSe(x, \hloc=0^+)
\end{align}
\end{subequations}
representing, respectively, the degeneracy-lifted entanglement entropy and
the reduction in entanglement entropy due to spontaneous symmetry breaking.
 
\begin{figure}[t]
\includegraphics[height=3.5in, angle=-90]{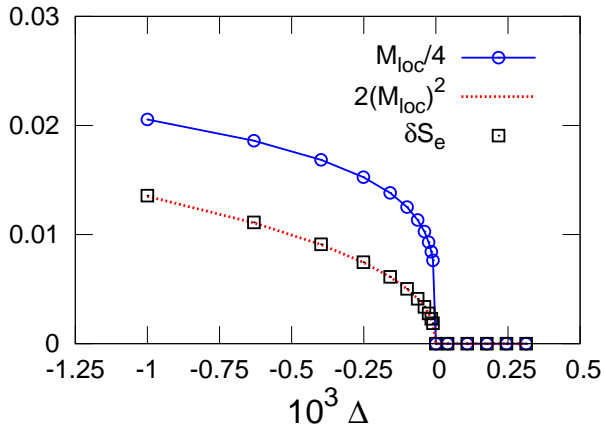}
\caption{\label{fig:M_loc-and-S_e:BFKM}(Color online)
Order parameter $\Mloc(\Delta,\hloc=0^+)$ and spontaneous-symmetry-breaking
part of the entanglement entropy $\dSe(\Delta)$ vs distance $\Delta=(J-J_c)/J_c$
from the quantum critical point of the Bose-Fermi Kondo model with an
Ising-symmetry coupling $K_0 g=7.5$ between an impurity spin $\Simp=\half$ and
a bosonic bath with exponent $s=0.8$.
On approach to the QCP from the local-moment phase ($\Delta<0$), $\dSe$
vanishes like $2 \Mloc^2$.}
\end{figure}

In Secs.\ \ref{sec:Kondo} and \ref{sec:Ander}, we report results for the
entanglement entropy in several quantum impurity Hamiltonians of the general
form
\begin{equation}
\label{H:gen}
H = H_{\host}+H_{\imp}+H_{\hostimp}
\end{equation}
where $H_{\host}$ describes one or more fermionic bands as well as, possibly, 
a bosonic bath. The term $H_{\imp}$ describes the isolated impurity, and
$H_{\hostimp}$ accounts for the coupling between the host and the impurity.
The fermionic bands are assumed to have a dispersion $\epsilon_{\bk}$ giving
rise to an idealized density of states
\begin{equation}
\label{rho_c}
\rho_c(\epsilon)
  = N_k^{-1} \sum_{\bk} \delta(\epsilon-\epsilon_{\bk})
  = \rho_0|\epsilon/D|^r \Theta(D-|\epsilon|),
\end{equation}
where $N_k$ is number of distinct $\bk$ values, $\Theta(x)$ is the Heaviside
function,  $D$ is the half-bandwidth, and $r$ is a band exponent. QCPs arise
in pure-fermionic Kondo and Anderson models in cases $0<r<1$ describing
pseudogapped hosts in which $\rho_c(\epsilon)$ vanishes in a sub-linear fashion
at the Fermi energy $\epsilon_F = 0$.

Another route to impurity QCPs is competition between fermionic and bosonic
environments for control of an impurity degree of freedom. Below we consider
models with a metallic fermionic density of states described by Eq.\
\eqref{rho_c} with $r=0$ as well as a bosonic bath assumed to have a dispersion
$\omega_{\bq}$ that gives rise to a density of states
\begin{equation}
\label{rho_phi}
\rho_{\phi}(\omega)
 = N_q^{-1} \sum_{\bq} \delta(\omega - \omega_{\bq})
 = K_0^2\omega_c^{1-s}|\omega|^s\Theta(\omega)\Theta(\omega_c-\omega),
\end{equation}
where $N_q$ is the number of distinct $\bq$ modes, $\omega_c$ is an
ultra-violet frequency cutoff, and $s$ is a bath exponent assumed to
lie in the range $1/2 < s < 1$.

In all of the models we consider it has been established that a QCP separates
a Kondo phase (corresponding to tuning parameter values $\Delta>0$) from a
local-moment phase (spanning $\Delta<0$), in which the Kondo effect is
destroyed. The models include ones in which the Kondo phase exhibits exact
screening, over-screening, or under-screening of the impurity spin. In each
case an appropriate order parameter for the quantum phase transition is the
$\hloc\to 0^+$ limit of the local magnetization $\Mloc(\Delta,\hloc) =
-\lim_{T\rightarrow 0}\langle \Simpz\rangle$, which vanishes throughout the
Kondo phase, and in the local-moment phase close to the QCP obeys
\begin{equation}
\label{beta:def}
\Mloc(\Delta, \hloc=0^+) \propto (-\Delta)^{\beta},
\end{equation}
where $\beta$ is the order-parameter exponent. At the critical value of the
tuning parameter, the local magnetization satisfies
\begin{equation}
\label{delta:def}
\Mloc(\Delta=0, \hloc) \propto |\hloc|^{1/\delta},
\end{equation}
where $\delta$ is another critical exponent.

We show very generally---independent of features such as particle-hole symmetry
or asymmetry, the degree of impurity charge fluctuations, whether the Kondo
phase involves exact, over-, or under-screening, and the presence or absence of
competition between fermionic bands and bosonic baths---that upon approach to
the QCP from the local-moment side, the entanglement entropy satisfies
\begin{equation}
\label{dS_e-vs-M_loc}
\dSe(\Delta, \hloc) = a \Mloc^2,
\end{equation}
where $a$ is a constant that depends on details of the model.
This connection is illustrated in Fig.\ \ref{fig:M_loc-and-S_e:BFKM},
which shows the variation of $\Mloc(\Delta,\hloc=0^+)$ and $\dSe(\Delta)$ with
reduced Kondo coupling $\Delta=(J-J_c)/J_c$ in the Ising-symmetry Bose-Fermi
Kondo model [defined in Eqs.\ \eqref{H:BFKM} below], for which we show that
$a=2$.

When combined with Eqs.\ \eqref{beta:def} and \eqref{delta:def}, Eq.\
\eqref{dS_e-vs-M_loc} implies that
\begin{subequations}
\begin{align}
\label{beta_e:def}
\dSe(\Delta<0, \hloc=0^+) \propto (-\Delta)^{\beta_e}, \\
\label{delta_e:def}
\dSe(\Delta=0, \hloc) \propto |\hloc|^{1/\delta_e}, 
\end{align}
\end{subequations}
with
\begin{subequations}
\begin{align}
\label{beta_e:scaling}
\beta_e = 2\beta, \\
\label{delta_e:scaling}
1/\delta_e = 2/\delta,
\end{align}
\end{subequations}
scaling relations that are demonstrated explicitly in the numerical results
presented below.

We solve the quantum impurity problems introduced above using the
NRG \cite{Wilson.75,Bulla.08} as adapted to treat quantum impurity problems
involving a pseudogapped fermionic density of states \cite{Bulla.97,Buxton.98}
and/or a sub-ohmic bath of bosons \cite{Bulla.03,Bulla.05,Glossop.05,Pixley.13}.
The impurity entanglement entropy is found via Eq.\ \eqref{S_e:imp} using
reduced density matrix eigenvalues $p_i$ expressed in terms of expectation
values (converged for large NRG iteration numbers corresponding to
asymptotically low temperatures) of certain impurity operators
specified in the sections that follow.
We use a Wilson discretization parameter $3\le\Lambda\le9$, a range that has
been shown previously to provide an accurate account of the critical
exponents \cite{Ingersent.02,Glossop.05,Glossop.07,Pixley.13}. Close to the QCP,
we find it necessary to employ quadruple-precision floating-point calculations
in order to accurately resolve the entanglement entropy, and in particular, the
value of $\dSe(\Delta)$ defined in Eq.\ \eqref{dS_e:def}.

\section{Kondo and Bose-Fermi Kondo Models}
\label{sec:Kondo}

The Kondo models under consideration are described by Hamiltonians of the
form of Eq.\ \eqref{H:gen} with
\begin{subequations}
\label{H:Kondo}
\begin{align}
\label{H:Kondo:host}
H_{\host}
& = \sum_{\bk,\sigma,\alpha} \epsilon_{\bk} \,
    c_{\bk\sigma\alpha}^{\dag} \, c_{\bk\sigma\alpha}^{\pdag}, \\
H_{\imp}
& = \hloc \Simpz, \\
H_{\hostimp}
& = J \, \mathbf{S}_{\imp} \cdot \sum_{\sigma,\sigma',\alpha}
    c_{0\sigma\alpha}^{\dag} \: \half \, \bm{\sigma}_{\sigma\sigma'} \:
    c_{0\sigma'\alpha}^{\pdag} \notag \\
& \quad {} + W \sum_{\sigma,\alpha}
    c_{0\sigma\alpha}^{\dag} \, c_{0\sigma\alpha}^{\pdag},
\end{align}
\end{subequations}
where $c_{\bk\sigma\alpha}$ destroys a conduction electron of wave vector
$\bk$, spin $z$ component $\sigma=\pm\half\equiv\;\uparrow,\downarrow$,
channel index $\alpha\in \{1, \ldots, K\}$, and energy $\epsilon_{\bk}$
satisfying Eq.\ \eqref{rho_c};
$c_{0\sigma\alpha}=N_k^{-1/2} \sum_{\bk} c_{\bk\sigma\alpha}$ destroys an
electron of spin $z$ component $\sigma$ and channel index $\alpha$ at the
impurity site; and $\mathbf{S}_{\imp}$ is the spin operator for a
spin-$\Simp$ impurity.
The host-impurity coupling is characterized by an antiferromagnetic exchange
$J>0$ and a potential scattering $W$.

In Sec.\ \ref{subsec:spin-half} we consider pseudogap Kondo models described
by Eqs. \eqref{H:gen} and \eqref{H:Kondo} with an impurity spin $\Simp=\half$
and channel numbers $K=1$ (exactly screened impurity) and $K=2$ (overscreened
impurity). We also consider the (one-channel, $\Simp=\half$) Ising-symmetry
Bose-Fermi Kondo model described by Eq.\ \eqref{H:gen} with
\begin{subequations}
\label{H:BFKM}
\begin{align}
H_{\host}
& = \sum_{\bk,\sigma,\alpha} \epsilon_{\bk} \,
    c_{\bk\sigma\alpha}^{\dag} \, c_{\bk\sigma\alpha}^{\pdag}
    + \sum_{\bq} \omega_{\bq} \, \phi_{\bq}^{\dag} \, \phi_{\bq}^{\pdag}, \\
H_{\imp}
& = \hloc \Simpz \\
H_{\hostimp}
& = J \, \mathbf{S}_{\imp} \cdot \sum_{\sigma,\sigma',\alpha}
    c_{0\sigma\alpha}^{\dag} \: \half \, \bm{\sigma}_{\sigma\sigma'} \:
    c_{0\sigma'\alpha}^{\pdag} \notag \\
{} + W \sum_{\sigma,\alpha} &
    c_{0\sigma\alpha}^{\dag} \, c_{0\sigma\alpha}^{\pdag}
    + g \Simpz \frac{1}{\sqrt{N_q}}
    \sum_{\bq} \bigl( \phi_{\bq}^{\dag} + \phi_{-\bq}^{\pdag} \bigl),
\end{align}
\end{subequations}
Here, a QCP may be present both for a metallic ($r=0$) and a semimetallic
($r>0$) conduction band. Section \ref{subsec:spin-one} treats a spin-one
impurity, focusing on the underscreened $K=1$ pseudogap model.
In each of these cases, we consider $W$ (and where present $g$) to be held
fixed and define the distance from criticality to be $\Delta = (J-J_c)/J_c$.

Even with the addition of a degeneracy-lifting field that couples to
$\Simpz$, the Kondo and Bose-Fermi Kondo Hamiltonians [Eqs.\
\eqref{H:Kondo} and \eqref{H:BFKM}] exhibit spin-rotation symmetry about
the $z$ axis and hence conserve the $z$ component of total spin. This ensures
that the eigenstates of $\rhoimp$ can be chosen to be the conventional basis
states $|m\rangle$ such that $\Simpz |m\rangle = m |m\rangle$. Then,
\begin{equation}
S_e = -\sum_{m = -S}^{S} p_m \ln p_m .
\end{equation}

\subsection{$\Simp=\half$ Kondo and Bose-Fermi Kondo models}
\label{subsec:spin-half}

For a spin-$1/2$ Kondo impurity and in the presence of spin-rotation
symmetry about the $z$ axis, the eigenvalues of $\rhoimp$ are just the
impurity spin-up and spin-down occupation probabilities
\begin{equation}
\label{eigen:spin-half}
p_{\pm 1/2} = \half \pm \Mloc,
\end{equation}
and Eq.\ \eqref{S_e:imp} reduces to
\begin{equation}
\label{S_e:spin-half:v1}
S_e = S_2(\half + \Mloc) ,
\end{equation}
where
\begin{equation}
S_2(x) = -x \ln x - (1-x) \ln(1-x)
\end{equation}
is the binary entropy function.
Expanding Eq.\ \eqref{S_e:spin-half:v1} for $|\Mloc|\ll \half$ gives
\begin{equation}
\label{S_e:spin-half}
S_e \simeq \ln 2 - 2 \Mloc^2 - (4/3) \Mloc^4,
\end{equation}
a result that holds for any $S=\half$ Kondo model, irrespective of the number
and dispersion of the conduction bands.

Equation \eqref{S_e:spin-half} implies that even in the presence of an
infinitesimal magnetic field, the entanglement entropy takes its maximum
possible value of $\ln 2$ at any magnetic QCP and throughout the Kondo phase.
This is true both when the impurity moment is exactly screened
with Fermi-liquid excitations (as is the case for $K=1$) and when it is
over-screened with a non-Fermi liquid many-body spectrum (as for $K\ge 2$).
Taking into account also Eq.\ \eqref{beta:def}, one finds that on approach to
the QCP from the local-moment side ($\Delta\to 0^-$),
\begin{equation}
\label{dS_e:spin-half}
\dSe(\Delta)\simeq 2 \Mloc^2\propto (-\Delta)^{2\beta},
\end{equation}
realizing Eq.\ \eqref{dS_e-vs-M_loc} with $a=2$ as well as
exemplifying Eqs.\ \eqref{beta_e:def} and \eqref{beta_e:scaling}.

\begin{figure}[t]
\includegraphics[width=2.2in,angle=-90]{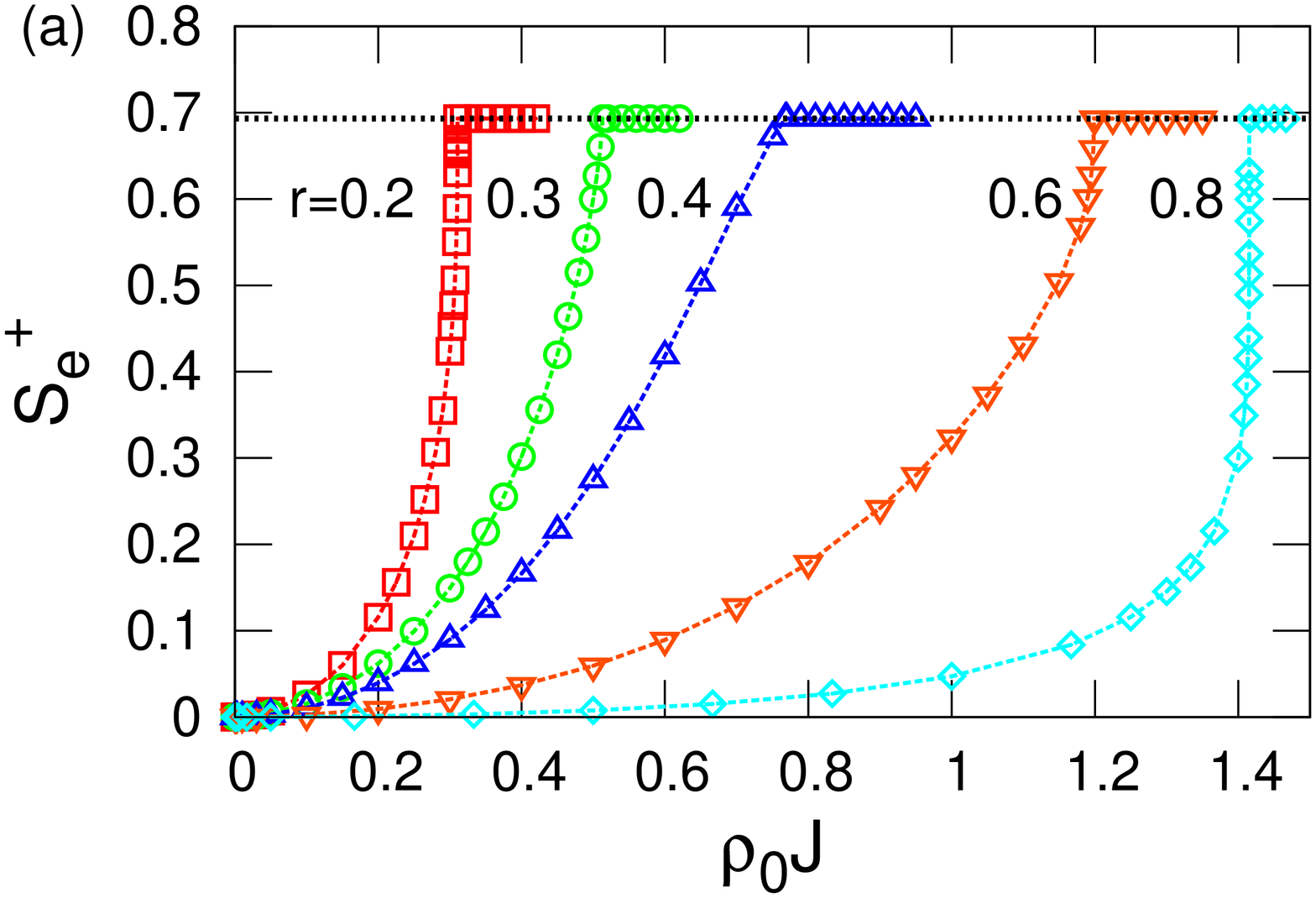}
\includegraphics[width=2.1in,angle=-90]{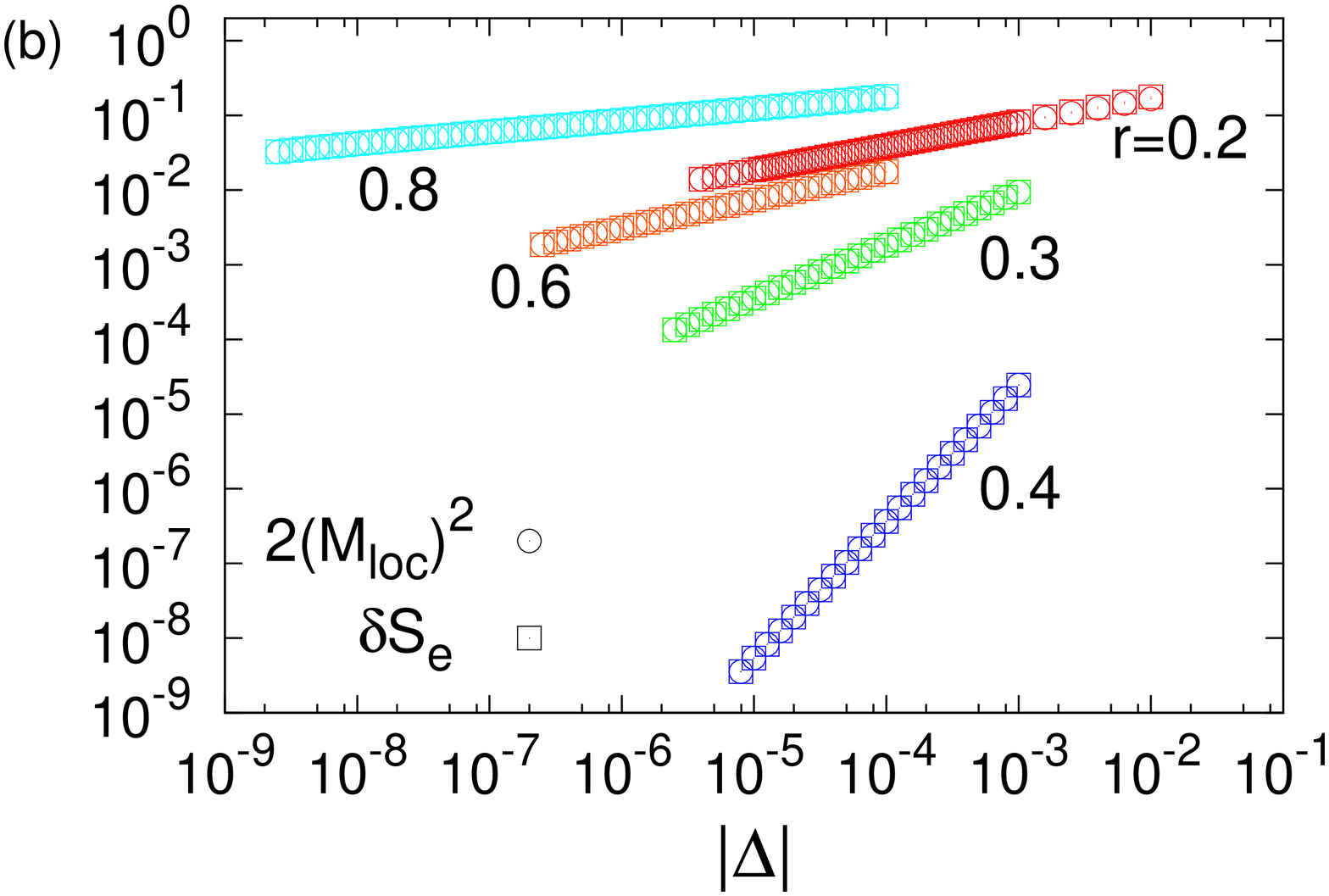}
\caption{\label{fig:S_e:spin-half}(Color online)
Entanglement entropy in the one-channel, $\Simp=\half$ pseudogap Kondo model
for various combinations of the band exponent $r$ and the dimensionless
potential scattering $\rho_0 W$.
(a) Degeneracy-lifted entanglement entropy $S_e^+$ vs dimensionless Kondo
coupling $\rho_0 J$. Results for $r=0.2$ and $r=0.3$ are at particle-hole
symmetry ($W=0$). Particle-hole-asymmetric results are shown for $r=0.4$
($\rho_0 W=0.109$), $r=0.6$ ($\rho_0 W=0.54$), and $r=0.8$ ($\rho_0 W=6.2$,
plotting $\rho_0 J / 6$ on the horizontal axis). In each case, $S_e^+$ takes
its maximum value of $\ln 2\simeq 0.693$ for all $J\ge J_c$.
(b) Spontaneous-symmetry-breaking part of the entanglement entropy
$\dSe(\Delta)$ and order-parameter $\Mloc$ (plotted as $2 \Mloc^2$)
vs distance $|\Delta|=(J_c-J)/J_c$ from the QCP in the local-moment phase
for several of the cases shown in (a). The coincidence of the two data sets
for each $r$ confirms Eq.\ \eqref{dS_e:spin-half} and straight-line fits
yield exponents listed in Table \ref{tab:spin-half}.}
\end{figure}

Another important conclusion that can be drawn from Eqs.\ \eqref{S_e:imp}
and \eqref{eigen:spin-half} is that $S_e^+$ vanishes only for
$|\Mloc|\to\half$, corresponding to a vanishing Kondo coupling $J$.
Even though the Kondo effect is destroyed for $0<J<J_c$ and the impurity
contributes $\ln 2$ to the zero-temperature entropy, the impurity remains
entangled with its environment and the ground state cannot be represented in
direct-product form.

We supplement these general theoretical considerations with NRG results for
several specific cases. For $\Simp=\half$, we only need to compute
$\Mloc = -\lim_{T\rightarrow 0}\langle \Simpz\rangle$, from which
$S_e$ can be obtained exactly via Eq.\ \eqref{S_e:spin-half:v1}.
Figure \ref{fig:S_e:spin-half}(a) shows the
degeneracy-lifted entanglement entropy $S_e^+$ vs $\rho_0 J$ for the
one-channel pseudogap Kondo model. This model is known \cite{Buxton.98} to have
(1) a particle-hole-symmetric Kondo-destruction QCP at $J = J_{c,s}(r)$, $W=0$
for any band exponent $r$ on the range $0<r<\half$, and
(2) a pair of particle-hole-asymmetric QCPs at $J = J_{c,a}(r)$, $W=\pm W_c(r)$
for any $r$ satisfying $r^*\simeq 3/8 < r < 1$.
The figure shows, both for the symmetric case $W=0$ (illustrated for $r=0.2$
and $0.3$) and the asymmetric one $W\ne 0$ (represented by data for $r=0.4$,
$0.6$, and $0.8$), that $S_e^+$ rises from 0 at $J=0$ to $\ln 2$
at $J=J_c$, and remains pinned at $\ln 2$ throughout the Kondo phase.

Figure \ref{fig:S_e:spin-half}(b) is a log-log plot of $2\Mloc^2$ and $\dSe$
vs $|\Delta|$ in the local-moment phase. Fitting to the power-law forms of
Eqs.\ \ref{beta:def} and \eqref{beta_e:def} yields the exponents $\beta$ and
$\beta_e$ listed in Table \ref{tab:spin-half}.
The values of $\beta$, along with ones of $1/\delta$ also listed, improve the
accuracy of those published previously \cite{Ingersent.02}.
For $r=0.3$, $0.4$, and $0.6$, the scaling Eq.\ \eqref{beta_e:scaling} is
obeyed to within the (small) estimated nonsystematic error of the exponents.
For $r=0.2$, $W=0$ and for $r=0.8$, $W\ne 0$ we find minor deviations from
scaling (at the $0.5\%$ and $1.5\%$ levels, respectively). In these cases, the
exponent $1/\delta$ takes small values that make it very difficult to reach
values of $|\Mloc|\ll\half$ and $\dSe\ll\ln 2$ even at $J=J_c$ when using the
smallest magnetic fields that we can apply numerically (of order
$\hloc = 10^{-36}D$ for quadruple-precision floating-point arithmetic).
This forces estimation of $\beta$ and $\beta_e$ at $J$ values further from $J_c$
where the asymptotic power-law form is obeyed less well.

For most values of $r$ the entanglement exponent satisfies $\beta_e<1$, meaning
that $dS_e^+/d\Delta$ diverges on approach to the QCP from the local-moment
side. However, for $0.35\lesssim r < \half$ at particle-hole symmetry, as well
as for $0.35\lesssim r \lesssim 0.45$ away from this symmetry (exemplified in
Table \ref{tab:spin-half} by $r=0.4$, $\rho_0 W=0.109$), one instead finds
$\beta_e>1$, leading to a much weaker feature in $S_e^+$ vs $\Delta$ at the
location of the quantum phase transition.
 
\begin{table}[t]
\begin{tabular}{l@{\extracolsep{2em}}l@{\extracolsep{1em}}%
                l@{\extracolsep{1.5em}}l@{\extracolsep{-2em}}d}
\multicolumn{1}{l}{$r$} &
  \multicolumn{1}{r}{$\rho_0 W$} &
    \multicolumn{1}{c}{$\beta$} &
      \multicolumn{1}{c}{$1/\delta$} &
        \multicolumn{1}{c}{$\qquad\beta_e$}
\\[.5ex] \hline \\[-2ex]
0.2 & 0     & 0.16025(1) & 0.02645(4) & 0.324(2) \\
0.3 & 0     & 0.35499(2) & 0.07398(2) & 0.71007(4) \\
0.4 & 0.109 & 0.57553(2) & 0.15606(3) & 1.15106(2) \\
0.6 & 0.54  & 0.18759(2) & 0.11696(5) & 0.3754(3) \\
0.8 & 6.2   & 0.07578(2) & 0.06373(3) & 0.156(2)
\end{tabular}
\caption{\label{tab:spin-half}
Exponents $\beta$, $\delta$, and $\beta_e$ defined in Eqs.\ \eqref{beta:def},
\eqref{delta:def}, and \eqref{beta_e:def}, respectively, for the $\Simp=\half$,
one-channel Kondo model with the five combinations of the band exponent $r$ and
dimensionless potential scattering $\rho_0 W$ illustrated in Fig.\
\ref{fig:S_e:spin-half}. Parentheses enclose the estimated nonsystematic error
in the last digit. The exponents obey Eq.\ \eqref{beta_e:scaling} within the
errors, apart from weak violations for $r=0.2$ and $r=0.8$, where the small
values of $1/\delta$ impede accurate evaluation of $\beta$ and $\beta_e$.}
\end{table}

We have also performed NRG calculations for the two-channel pseudogap Kondo
model [the case of Eq.\ \eqref{H:Kondo} with $K=2$], which is
predicted \cite{Buxton.98,Schneider.11,2-chan:note} for $0<r<r_{\max}$ (where
$0.23<r_{\max}<0.24$) to have a particle-hole-symmetric QCP between a
local-moment phase and a non-Fermi liquid overscreened-Kondo phase. Equation
\eqref{dS_e:spin-half} is predicted to hold independent of $K$, and indeed our
numerical results summarized in Table \ref{tab:two-chan} support this picture:
any differences between the cases $K=1$ and $K=2$ arise just from the two models
having their own order-parameter exponents $\beta(r)$.

\begin{table}[t]
\begin{tabular}{l@{\extracolsep{1em}}l@{\extracolsep{1em}}%
                l@{\extracolsep{-1em}}d}
\multicolumn{1}{c}{$r$} &
  \multicolumn{1}{c}{$\beta$} &
    \multicolumn{1}{c}{$1/\delta$} &
      \multicolumn{1}{c}{$\qquad\beta_e$}
\\[.5ex] \hline \\[-2ex]
0.05 & 0.050(1) & 0.001815(2) & 0.140(5) \\
0.1  & 0.150(2) & 0.013075(2) & 0.327(4) \\
0.15 & 0.330(1) & 0.036233(8) & 0.667(3) \\
0.2  & 0.908(3) & 0.09205(2)  & 1.815(9)
\end{tabular}
\caption{\label{tab:two-chan}
Exponents $\beta$, $1/\delta$, and $\beta_e$ defined in Eqs.\ \eqref{beta:def},
\eqref{delta:def}, and \eqref{beta_e:def}, respectively, for the $\Simp=\half$,
two-channel Kondo model at particle-symmetry ($W=0$) and four different values
of the band exponent $r$. Parentheses enclose the estimated nonsystematic error
in the last digit. The exponents obey Eq.\ \eqref{beta_e:scaling} to within a
margin that is within nonsystematic errors for $r=0.2$, but becomes larger for
lower values of $r$, likely due to the small values of exponent $1/\delta$.}
\end{table}

The final $\Simp=\half$ model that we have studied numerically is the
Ising-symmetry Bose-Fermi Kondo model [Eq.\ \eqref{H:BFKM}]. As exemplified
in Fig.\ \ref{fig:M_loc-and-S_e:BFKM} for the case of conduction-band exponent
$r=0$ and a sub-Ohmic bosonic-bath exponent $s=0.8$, NRG calculations once
again confirm that $\dSe$ obeys Eq.\ \eqref{dS_e:spin-half} on approach to
the Kondo-destruction QCP from the local-moment side ($\Delta<0$), whereas in
the Kondo-screened phase ($\Delta>0$) we find $S_e(\Delta,0)=\ln 2$. Just as
in pure-fermionic Kondo models, $S_e^+$ rises on approach to phase
boundary from the local-moment side to attain is maximum possible value at the
QCP and throughout the Kondo phase.

The behavior of the entanglement entropy in the Bose-Fermi Kondo model is in
marked contrast with that of the sub-Ohmic spin-boson model, described by
Eq.\ \eqref{H:gen} with
\begin{subequations}
\label{H:SBM}
\begin{align}
H_{\host}
& = \sum_{\bq} \omega_{\bq} \, \phi_{\bq}^{\dag} \, \phi_{\bq}^{\pdag}, \\
H_{\imp}
& = \hloc \Simpz - \Delta_x \Simp^x \\
H_{\hostimp}
& = g \Simpz \sum_{\bq} \bigl( \phi_{\bq}^{\dag} + \phi_{-\bq}^{\pdag} \bigl),
\end{align}
\end{subequations}
where the dispersion $\omega_{\bq}$ satisfies Eq.\ \eqref{rho_phi} with
$0<s<1$.
NRG studies \cite{LeHur.07,LeHur.08} show that $S_e$ rises on approach
to the QCP from \textit{either} phase, and exhibits a cusp peak at
a non-universal value that falls short of the maximum possible value $\ln 2$.
This departure from the entanglement found in the Bose-Fermi Kondo model seems
surprising, given that both models map \cite{Spohn.85,Kirchner.08} to a
classical Ising model with long-range ferromagnetic interactions that decay
for large separations $d$ as $1/d^{1+s}$ where $s$ is the bosonic bath
exponent. The two quantum-mechanical models are thought to share the
same critical exponents \cite{Zhu.02,Glossop.05,Glossop.07}, at least for
$\half<s<1$.

However, the classical mapping and the
resulting critical exponents describing the response to a local magnetic
field along the $z$ axis take no account of the different global
symmetries of the two models.
The Kondo model exhibits a global $U(1)$ spin symmetry
leading to a conserved total spin $z$ component, as well as an emergent
$SU(2)$ spin symmetry at the Kondo-screened fixed point. By contrast, the
spin-boson model exhibits no spin symmetry at the Hamiltonian level (at least
in nontrivial cases where $\Delta_x$ and $g$ are both nonzero), while
the two stable fixed points exhibit emergent $U(1)$ symmetries leading to
conservation of total spin $x$ at the delocalized fixed point (analogous the
the Kondo phase) and conservation of total spin $z$ at the localized
(local-moment) fixed point.

In the absence of a globally conserved total spin $z$ component we cannot
assume that the eigenstates of the impurity reduced density operator are
also eigenstates of $\Simp^z$. Instead, the properties
$\mathrm{Tr}\,\rhoimp=1$ and
$\mathrm{Tr}(\rhoimp\bm{\sigma}) = 2\langle\mathbf{S}\rangle$
are sufficient to fully specify the impurity reduced density operator as
\begin{equation}
\rhoimp = \half \, I + \langle\mathbf{S}_{\imp}\rangle \cdot \bm{\sigma},
\end{equation}
where $I$ is the $2\times 2$ identity operator. Then the entanglement
entropy is given exactly by Eq.\ \eqref{S_e:imp} with two eigenvalues
\cite{LeHur.08}
\begin{equation}
\label{eigen:general}
p_{\pm} = \half \pm \sqrt{\langle\Simp^x\rangle^2 + \langle\Simp^y\rangle^2
  + \langle\Simpz\rangle^2}.
\end{equation}
Clearly, $S_e$ attains its maximum possible value of $\ln 2$ if and only if
$\langle\mathbf{S}_{\imp}\rangle = \mathbf{0}$, a condition that is satisfied
throughout the Kondo phase of the Bose-Fermi Kondo model and also at the QCP.
By contrast, the nonzero value of $\Delta_x$ necessary to produce a quantum
phase transition in the sub-Ohmic spin-boson model leads to a nonzero
$\langle\Simp^x\rangle$ in both phases. We expect that for fixed $g$,
$S_e(\hloc=0)$ decreases smoothly from $\ln 2$ as $\Delta_x$ increases from
zero, and $S_e(\hloc=0)$ shows no feature at the QCP. The addition of
an infinitesimal field along the $z$ axis creates a nonzero
$\langle\Simpz\rangle$ only in the delocalized phase, thereby
yielding \cite{LeHur:note} a correction $\dSe$ obeying Eqs.\
\eqref{dS_e-vs-M_loc}, \eqref{beta_e:def}, and \eqref{beta_e:scaling}, and
leading $S_e^+$ to exhibit a cusp peak in all cases where $2\beta < 1$. We
speculate that similar behavior would arise in the Ising-symmetry Bose-Fermi
Kondo model in the presence of a transverse local field \cite{Nica.13},
although we have not explicitly tested this conjecture.

\subsection{$\Simp=1$ single-channel pseudogap Kondo model}
\label{subsec:spin-one}

For a spin-1 Kondo impurity and in the presence of spin-rotation symmetry
about the $z$ axis, the eigenvalues of $\rhoimp$ can be parameterized
\begin{subequations}
\label{eigen:spin-one}
\begin{align}
p_{\pm 1} &= \half(n_0\pm \Mloc), \\
p_0 &= 1-n_0,
\end{align}
\end{subequations}
where we have introduced $n_0=\langle (\Simpz)^2 \rangle$. For such an
impurity, we focus exclusively on the one-channel pseudogap Kondo model.
At particle-hole symmetry ($W=0$), this model has a QCP for any band
exponent $r$ on the range $0<r<r_{\max}$ where $0.26<r_{\max}<0.27$, while
for $W\ne0$ there is a pair of asymmetric QCPs for any $r^*\simeq 0.245<r<1$
\cite{Buxton.98}. Each QCP separates (1) a local-moment phase
with a decoupled spin-one impurity degree of freedom and a ground-state
moment-squared $\mu^2=\Simp(\Simp+1)=2$ from (2) an under-screened Kondo
phase in which the impurity spin and the conduction band form a many-body
ground state with a residual moment-squared $\mu^2=3(2+r)/8$ for $W=0$ or
$\mu^2=3/4$ for $W\ne 0$. 

In the Kondo phase, the system shows an infinitesimal response to an
infinitesimal degeneracy-lifting field, and we can take the eigenvalues
of $\rhoimp$ to be $p_{\pm 1}=p_0=1/3$, which corresponds to $\Mloc=0$,
$n_0=2/3$, and $S_e(\Delta,0)=\ln 3$. Interestingly, even though the spin-$1$
impurity is only partially screened, the entanglement entropy still takes its
maximal value of $\ln(2\Simp+1)$.

In the local-moment phase, by contrast, the ground state exhibits
spontaneously broken $SU(2)$ symmetry, and it is natural that $n_0$
should rise above $2/3$. However, it is plausible (and we confirm below)
that close to the QCP,
deviations of $n_0$ from $2/3$ will be smaller than those of the order
parameter $\Mloc$ from zero, and that the former can safely be neglected.
More specifically, one can conjecture that
\begin{equation}
\label{beta_n:def}
n_0(\hloc=0^+) - 2/3 \propto (-\Delta)^{\beta_n} ,
\end{equation}
with $\beta_n>\beta$. Substituting Eqs.\ \eqref{eigen:spin-one} into
Eq.\ \eqref{S_e:imp}, setting $n_0=2/3$ and keeping only leading terms
in $\Mloc$, one arrives at the result
\begin{equation}
\label{dS_e:spin-one}
\dSe(\Delta) \approx \frac{3}{4}\Mloc^2 \propto (-\Delta)^{2\beta},
\end{equation}
providing a realization of Eq.\ \eqref{dS_e-vs-M_loc} with $a=3/4$.
As was the case for $\Simp=\half$, the predicted behavior is consistent
with Eqs.\ \eqref{beta_e:def} and \eqref{beta_e:scaling}.

\begin{figure}[t]
\hspace*{2em}\includegraphics[width=2.1in,angle=-90]{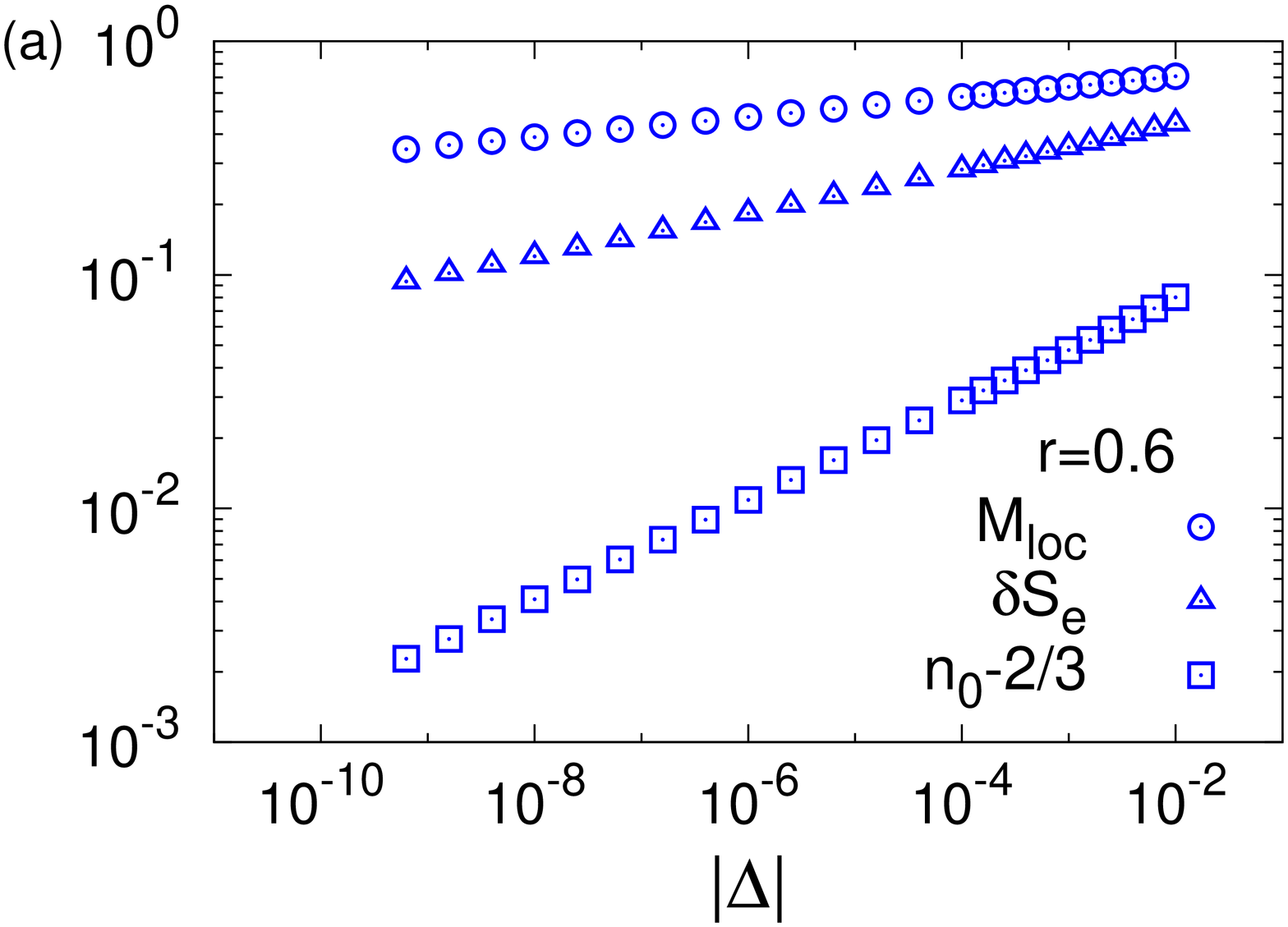}
\includegraphics[width=2.3in,angle=-90]{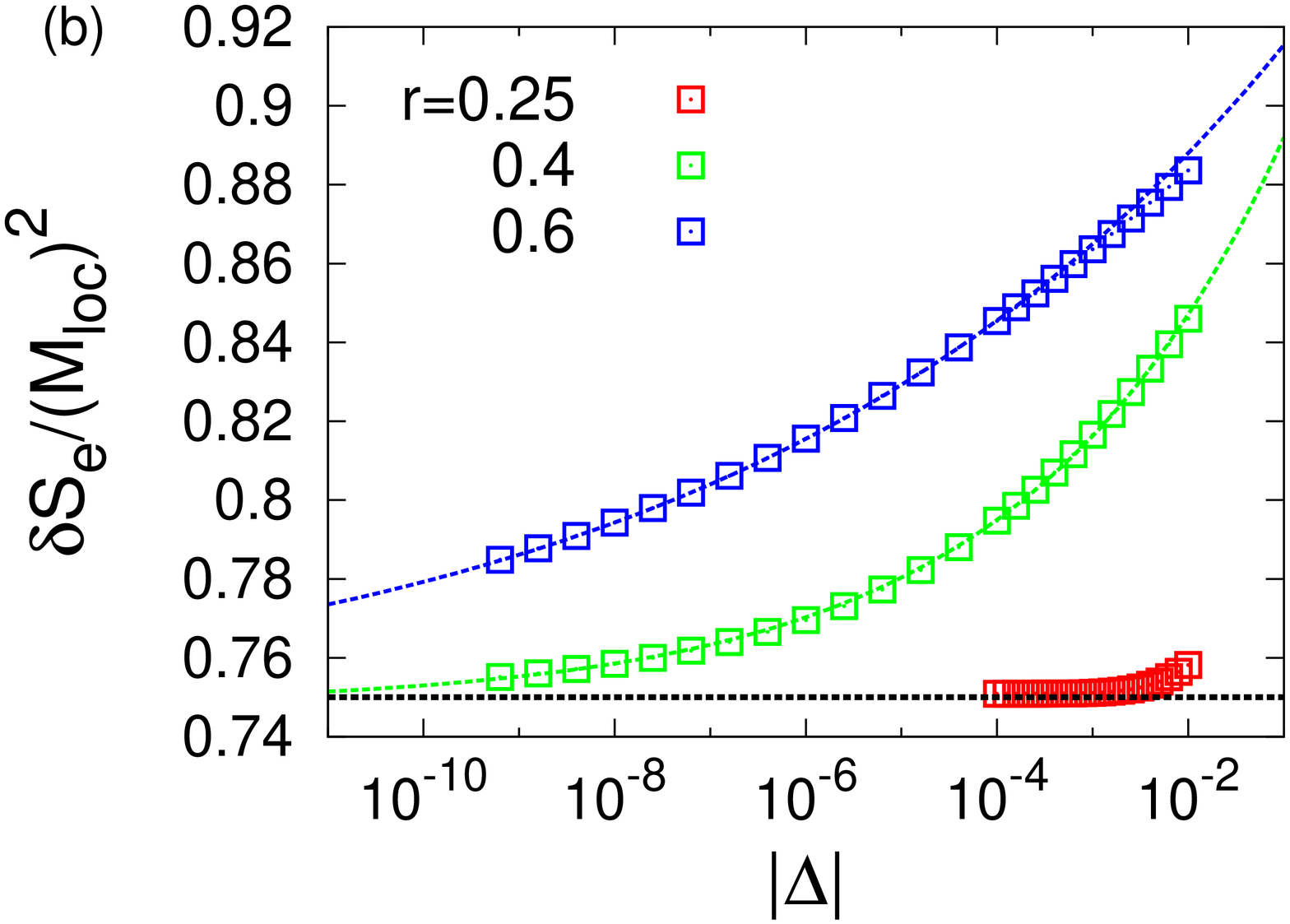}
\caption{\label{fig:S_e:spin-one}(Color online)
Approach from the local-moment side to the quantum critical point of the
one-channel, $\Simp=1$ pseudogap Kondo model for three combinations of
the pseudogap exponent and the dimensionless potential scattering:
$(r,\,\rho_0 W)=(0.25,\, 0)$, $(0.4,\,0.5)$, and $(0.6,\,0.8)$.
(a) Log-log plot of $\Mloc$, $n_0-2/3$, and $\dSe$ vs $|\Delta|$ for $r=0.6$,
with the linear variations exemplifying Eqs.\ \eqref{beta:def},
\eqref{beta_n:def}, and \eqref{beta_e:def}, respectively.
(b) Ratio $\dSe/\Mloc^2$ vs $|\Delta|$.
The convergence of the ratio to $0.75$ as $\Delta\to 0^-$
(albeit at a rate that decreases with increasing $r$) confirms the
validity of Eq.\ \eqref{dS_e:spin-one}.}
\end{figure}

Figure \ref{fig:S_e:spin-one} presents NRG results for three combinations of
the pseudogap exponent and the dimensionless potential scattering:
$(r,\,\rho_0 W)=(0.25,\, 0)$, $(0.4,\,0.5)$, and $(0.6,\,0.8)$.
Here, we have computed $\Mloc=-\langle\Simpz\rangle$ and
$n_0=\langle(\Simpz)^2\rangle$, then used Eqs.\ \eqref{S_e:imp}
and \eqref{eigen:spin-one} to find $S_e$.
Figure \ref{fig:S_e:spin-one}(a) shows that on approach to each QCP from the
local-moment side, $\Mloc$, $n_0-2/3$, and $\dSe$ vanish in power-law
fashion according to Eqs.\ \eqref{beta:def}, \eqref{beta_n:def}, and
\eqref{beta_e:def}, respectively. The exponents extracted from plots such
as this are summarized in Table \ref{tab:spin-one} for the three examples
shown in the figure, plus the particle-hole-symmetric QCPs for $r=0.1$,
$0.15$, and $0.2$. In each case, the exponents satisfy $\beta_n > 3 \beta$,
confirming the conjecture that deviations of $n_0$ from $2/3$ can be neglected
in the vicinity of the QCP. Equation \eqref{beta_e:scaling} is also satisfied
with one minor violation (for $r=0.25$) and one more significant one for
$r=0.6$ that can be attributed to the same cause (a small exponent $1/\delta$)
as in the one-channel $\Simp=\half$ Kondo model.

Figure \ref{fig:S_e:spin-one}(b) plots the ratio $\dSe/\Mloc^2$ on the
same logarithmic $|\Delta|$ scale as used for Figure \ref{fig:S_e:spin-one}(a).
For $r=0.25,\; W=0$, the ratio converges rapidly to $0.75$, as predicted in Eq.\
\eqref{dS_e:spin-one}. The convergence is rather slower for $r=0.4,\;\rho_0 W=0.5$,
and markedly slower for $r=0.6,\;\rho_0 W=0.8$. This trend can be explained by
the taking into account the leading correction to Eq.\ \eqref{dS_e:spin-one} in
situations where $\beta_n > 2\beta$, namely,
\begin{equation}
\dSe(\Delta) \approx \frac{3}{4}\Mloc^2 + \frac{9}{32}\Mloc^4,
\end{equation}
and noting in Table \ref{tab:spin-one} the decrease in the value of $\beta$ as
$r$ increases toward 1. We can extrapolate the ratio $\dSe/\Mloc^2$ for
$\Delta\rightarrow 0$ to obtain $0.7499(3)$ and $0.746(6)$ for $r=0.4$ and
$r=0.6$, respectively, in excellent agreement with the predicted
value $a=3/4$. We therefore take Figure \ref{fig:S_e:spin-one}(b) as
providing confirmation of the relation Eq.\ \eqref{dS_e:spin-one}. 

\begin{table}[t]
\begin{tabular}{l@{\extracolsep{2em}}l@{\extracolsep{1em}}%
                l@{\extracolsep{-1em}}d@{\extracolsep{2em}}l}
\multicolumn{1}{c}{$r$} &
  \multicolumn{1}{c}{$\rho_0 W$} &
    \multicolumn{1}{c}{$\beta$} &
      \multicolumn{1}{c}{$\qquad\beta_n$} &
        \multicolumn{1}{c}{$\beta_e$}
\\[.5ex] \hline \\[-2ex]
0.1  & 0   & 0.062(5)   & 0.19(1)   & 0.13(1) \\
0.15 & 0   & 0.116(1)   & 0.386(5)  & 0.24(1) \\
0.2  & 0   & 0.207(1)   & 0.787(4)  & 0.423(4) \\
0.25 & 0   & 0.51(1)    & 2.7(1)    & 1.00(5) \\
0.4  & 0.5 & 0.08960(5) & 0.4509(3) & 0.180(1) \\
0.6  & 0.8 & 0.04300(5) & 0.2128(2) & 0.089(1)
\end{tabular}
\caption{\label{tab:spin-one}
Exponents $\beta$, $\beta_n$, and $\beta_e$ defined in Eqs.\ \eqref{beta:def},
\eqref{beta_n:def}, and \eqref{beta_e:def}, respectively, for the $\Simp=1$,
one-channel Kondo model six combinations of the band exponent $r$ and the
dimensionless potential scattering $\rho_0 W$. Parentheses enclose the
estimated nonsystematic error in the last digit. In each case,
$\beta_n\ge 3\beta$, implying that variations of $n_0$ are negligible
compared to those of $\Mloc$ in the vicinity of the quantum critical point.
The exponents obey Eq.\ \eqref{beta_e:scaling} within the errors, apart from a
weak violation for $r=0.2$ and a stronger one for $r=0.6$, where the small
value $1/\delta=0.02589(2)$ impedes accurate evaluation of $\beta$
and $\beta_e$.}
\end{table}

\section{Anderson Models}
\label{sec:Ander}

In this section we consider non-degenerate Anderson impurity models
characterized by an impurity Hamiltonian
\begin{equation}
\label{H:Ander:imp}
H_{\imp} = \epsilon_d \, n_d+ U n_{d\uparrow} \, n_{d\downarrow}
  + \hloc \bigl( n_{d\uparrow} - n_{d\downarrow} \bigr)/2,
\end{equation}
where $d_{\sigma}$ destroys an electron of energy $\epsilon_d$ and spin $z$
component $\sigma$ electron at the impurity site, $n_{d\sigma} =
d_{\sigma}^{\dag} d_{\sigma}^{\pdag}$ and $n_d=n_{d\uparrow}+n_{d\downarrow}$
are impurity number operators, and $U$ is the on-site Coulomb interaction. Such
models differ from their Kondo counterparts by allowing charge fluctuations on
the impurity site. We will show that such fluctuations can significantly modify
the behavior of the entanglement entropy in the vicinity of the continuous
quantum phase transition between Kondo-screened and local-moment
(Kondo-destroyed) phases. In particular, in the presence of an infinitesimal
symmetry-breaking field, the entanglement entropy in general neither exhibits
a peak at the QCP (as it does in the sub-Ohmic spin-boson model) nor attains
its maximum possible value at the QCP and throughout the Kondo phase (as is
the case in the Kondo and Bose-Fermi Kondo models discussed in Sec.\
\ref{sec:Kondo}).

Under conditions of spin-rotation symmetry about the $z$ axis, the eigenstates
of the impurity reduced density operator can be taken \cite{Larsson.05}
to be the conventional basis states $|0\rangle$,
$|\sigma\rangle=d_{\sigma}^{\dag}|0\rangle$, and
$|2\rangle=d_{\uparrow}^{\dag}d_{\downarrow}^{\dag}|0\rangle$, with
eigenvalues that can be written as
\begin{subequations}
\label{eigen:Ander}
\begin{align}
p_2 &= \langle n_{d\uparrow} \, n_{d\downarrow} \rangle, \\
p_{\uparrow} &=  f/2 + \Mloc, \\
p_{\downarrow} &= f/2 - \Mloc, \\
p_0 &= 1 - f - p_2,
\end{align}
\end{subequations}
in terms of the local-moment fraction (i.e., single-occupation probability)
$f=\langle n_d \rangle - 2p_2$ and the local moment
$\Mloc = \langle\half(n_{d\uparrow}-n_{d\downarrow})\rangle$,
such that $0\le f \le 1$ and $|\Mloc|\le f/2$. Then Eq.\ \eqref{S_e:imp} can
be written
\begin{equation}
S_e = f S_2 \biggl(\frac{1}{2} + \frac{M_{\loc}}{f} \biggr)
+ (1-f) S_2 \biggl( \frac{p_2}{1-f} \biggr) ,
\end{equation}
which can be interpreted as the sum of a binary spin entanglement entropy
with weight $f$ and a binary charge entanglement entropy with weight $(1-f)$.
The parallel between the spin and charge parts of $S_e$ can be made clearer by
defining a ``local charge'' $\Qloc = p_2 - p_0$, satisfying
$|\Qloc|\le 1-f$, so that
\begin{equation}
\frac{p_2}{1-f}  = \frac{1}{2} + \frac{\Qloc}{2(1-f)}.
\end{equation}

In the absence of a symmetry-breaking field $\hloc$, we can set $\Mloc=0$
in Eqs.\ \eqref{eigen:Ander}, and use Eq. \eqref{S_e:imp} to obtain
\begin{align}
\label{S_e:Ander:h=0}
S_e(\hloc=0) =
&-(1-f-p_2)\ln(1-f-p_2) \notag \\
&- f\ln(f/2) - p_2\ln p_2.
\end{align}
Since $f$ and $p_2$ are nontrivial
functions of $U$, $\epsilon_d$, and $\Gamma$, it is clear that Eq.\
\eqref{S_e:Ander:h=0} can encompass much richer behavior than the
corresponding Kondo-model result $S_e(\hloc=0) = \ln 2$.
Differentiating Eq. \eqref{S_e:Ander:h=0} with respect to $p_2$ for a
fixed local-moment fraction $f$ yields
\begin{equation}
\label{S_e:slope}
\left( \frac{\partial S_e}{\partial p_2} \right)_{f}
  = \ln \frac{p_0}{p_2} ,
\end{equation}
which implies that for a given value of $f$, $S_e(\hloc=0)$ is greatest
for equal occupation of the impurity configurations $|0\rangle$ and
$|2\rangle$ (i.e., $\Qloc=0$), and is smallest when one or other of the
configurations is ruled out.

In the local-moment phase near the boundary with the Kondo phase, we expect
an infinitesimal field $\hloc=0^+$ to establish a local magnetization
$|\Mloc|\ll\half$ with negligible shift of $f$ and $p_2$. Under these
circumstances, an expansion of $S_e$ in powers of $\Mloc$ shows the
spontaneous-symmetry-breaking part of the entanglement entropy to be
\begin{equation}
\label{dS_e:Ander}
\dSe(\Delta) \simeq 2 \Mloc^2/f \propto ( - \Delta)^{2\beta},
\end{equation}
providing yet another realization of Eqs.\ \eqref{dS_e-vs-M_loc},
\eqref{beta_e:def}, and \eqref{beta_e:scaling}, this time with $a=2/f$.
Since the local-moment fraction satisfies $0\le f\le 1$, the critical
part of the entanglement in nondegenerate Anderson models is generally
enhanced by a factor of $1/f$ compared to its counterpart in
$\Simp=\half$ Kondo models [see Eq.\ \eqref{dS_e:spin-half}].

For the purposes of numerical study, we focus on the one-channel Anderson
impurity model described by the Hamiltonian in Eq.\ \eqref{H:gen} with
$H_{\host}$ as given in Eq.\ \eqref{H:Kondo:host}, $H_{\imp}$ as in
Eq.\ \eqref{H:Ander:imp} and
\begin{equation}
H_{\hostimp} = \frac{V}{\sqrt{N_k}}
\sum_{\textbf{k},\sigma} \bigl(c_{\textbf{k} \sigma}^{\dag}
   d_{\sigma}^{\pdag} + \mathrm{H.c.} \bigr),
\end{equation}
The hybridization matrix element $V$ between the impurity site and the
conduction band is conventionally re-expressed in terms of the
hybridization width $\Gamma = \pi \rho_0 V^2$.
We have computed $\langle n_{d,\sigma} \rangle$ and
$\langle n_{d,\uparrow} n_{d,\downarrow} \rangle$, then used Eqs.\
\eqref{S_e:imp} and \eqref{eigen:Ander} to find $S_e$.

In the following we take $U$ and $\epsilon_d$ to be fixed, either at
particle-hole symmetry ($\epsilon_d=-U/2$) or away from it
($\epsilon_d\ne-U/2$). We then find the location of the QCP at a critical
hybridization width $\Gamma_c(U,\,\epsilon_d)$, and thereafter define the
distance from criticality as $\Delta=(\Gamma-\Gamma_c)/\Gamma_c$. The
critical responses to a local magnetic field near the symmetric and
asymmetric QCPs of the pseudogap Anderson model belong in the same
universality classes as the respective QCPs of the $\Simp=\half$ Kondo
model \cite{Buxton.98,Ingersent.02,Fritz.04,Glossop.11,Pixley.12}.

In the subsections that follow, we first consider two special cases
($U=-2\epsilon_d$ and $U=\infty$) in which $S_e(\hloc=0)$ in
Eq.\ \eqref{S_e:Ander:h=0} reduces to a function of one variable,
the local-moment fraction, thereby simplifying analysis of the
behavior of the entanglement entropy in the vicinity of the QCP.
Afterward, we present illustrative examples of the entanglement
properties for more general cases.

\begin{figure}[h!]
\includegraphics[width=2.2in,angle=-90]{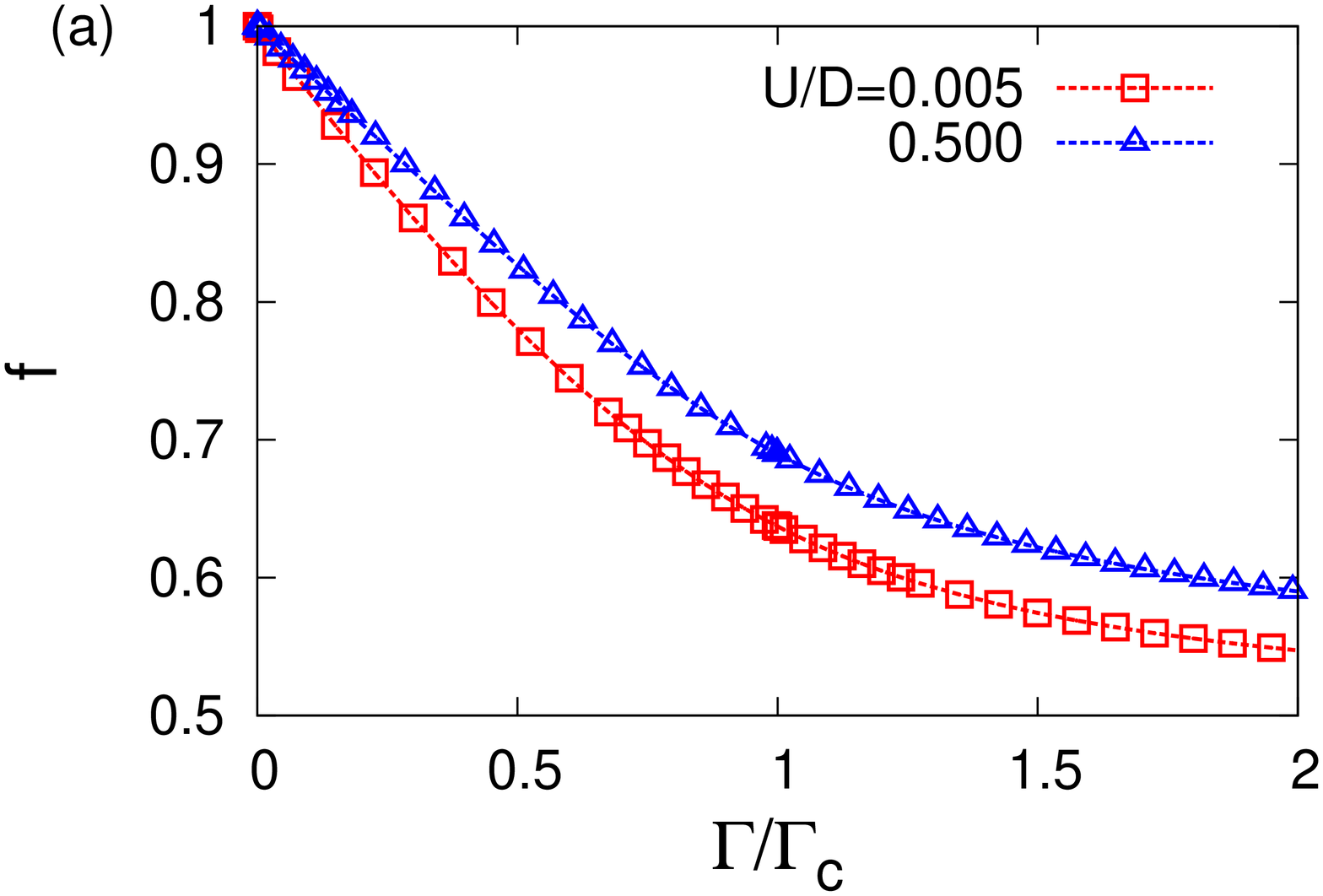} \\[-3ex]
\includegraphics[width=2.2in,angle=-90]{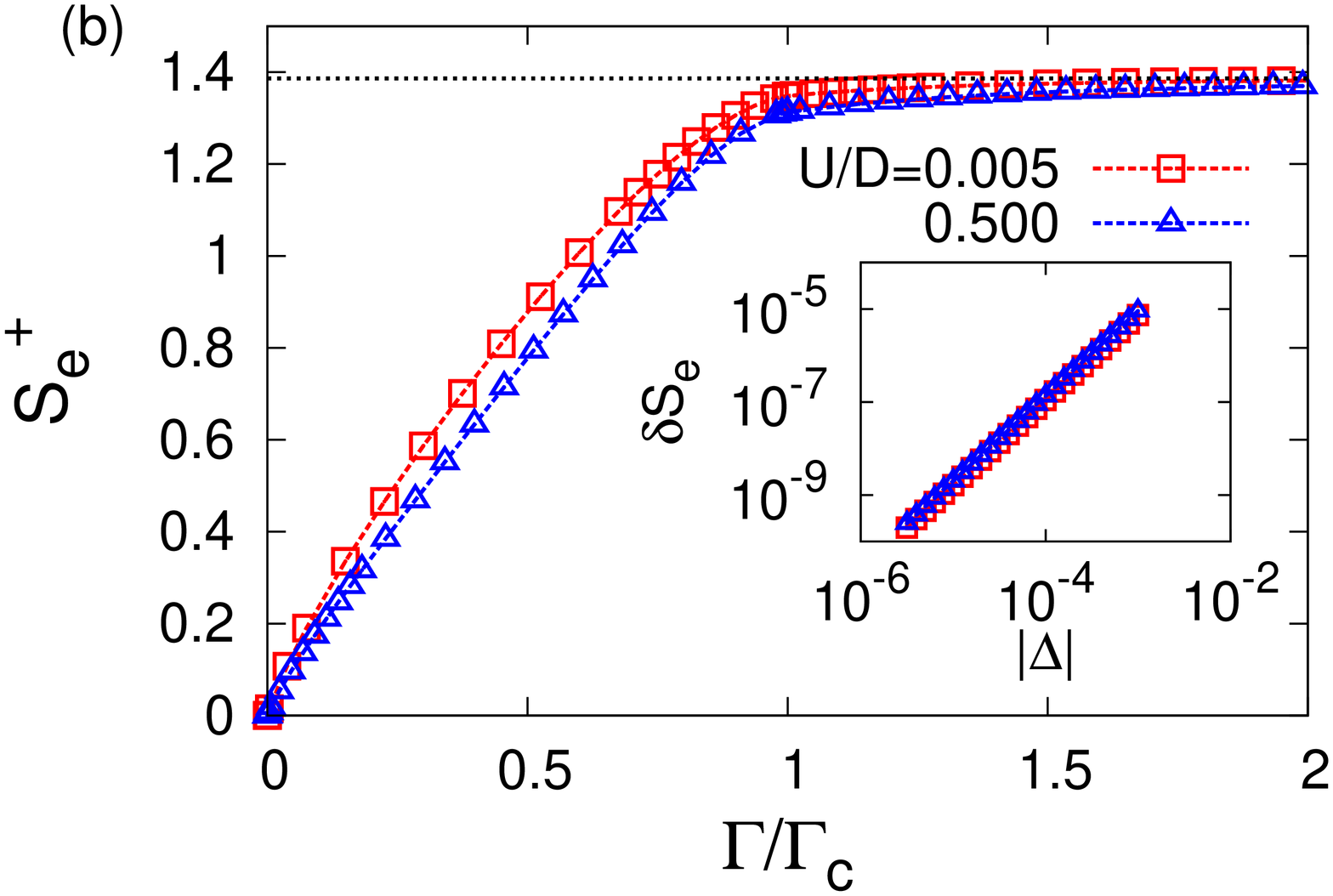} \\[-3ex]
\hspace{1ex}\includegraphics[width=2.15in,angle=-90]{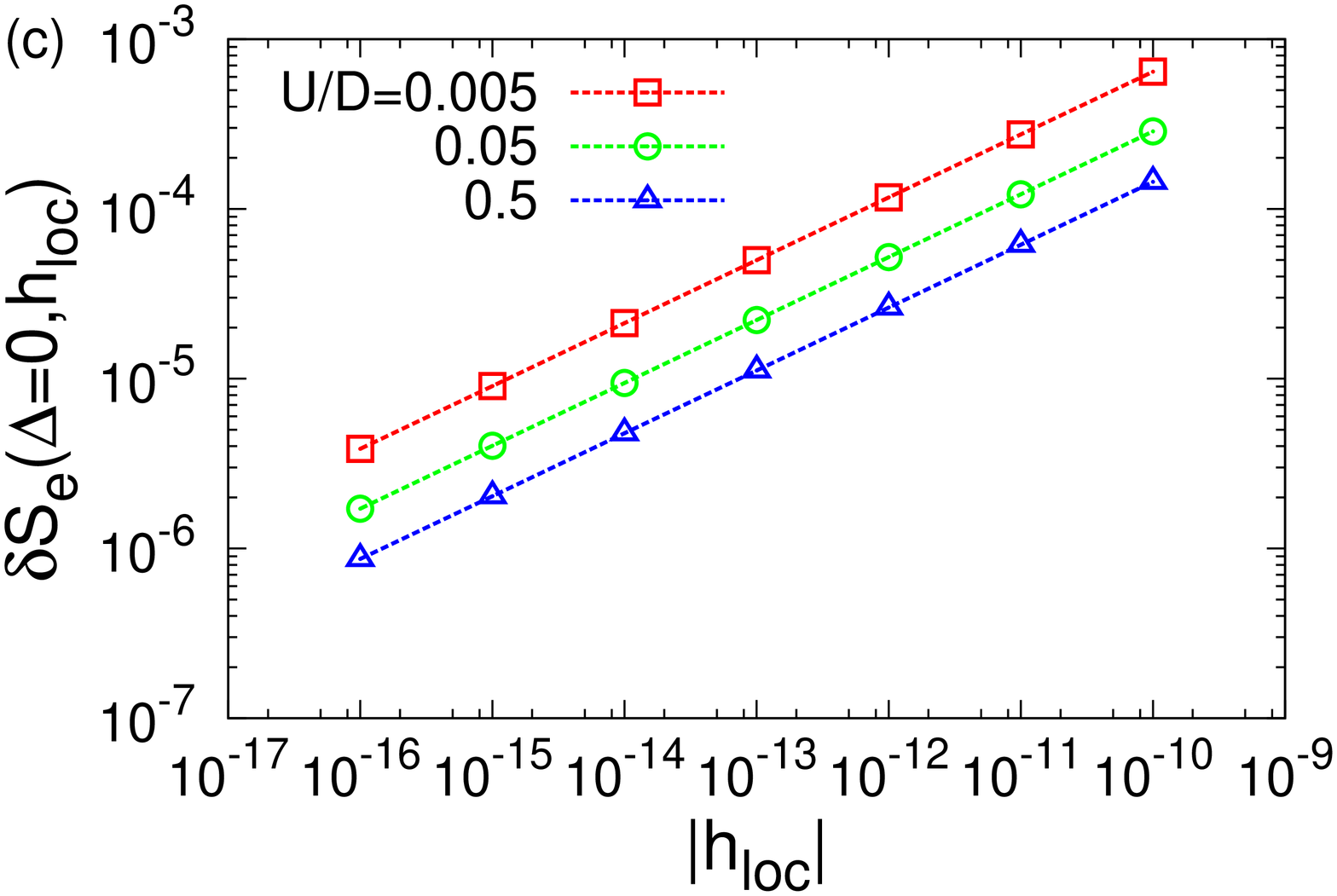}
\caption{\label{fig:Ander:symm}(Color online)
Particle-hole-symmetric pseudogap Anderson model with band exponent
$r=0.4$ for different values of $U/D$:
(a) Local-moment fraction $f$ and (b) symmetry-lifted entanglement
entropy $S_e^+$, both vs hybridization width $\Gamma$
scaled by its critical value $\Gamma_c$. In the limit of large $\Gamma$,
$S_e^+$ approaches its maximum value of $\ln 4\approx 1.386$.
Inset to (b): Spontaneous-symmetry-breaking part of the entanglement
entropy $\dSe$ vs $|\Delta|$ [where $\Delta=(\Gamma-\Gamma_c)/\Gamma_c$]
for the two cases shown in (a) and (b). The linear variations of the
points on this log-log plot are consistent with Eq.\ \eqref{beta_e:def}
with $\beta_e=1.8288(1)$.
(c) Local-field-dependent part of the entanglement entropy
$\dSe$ vs $\hloc$ at the critical hybridization width, showing
behavior consistent with Eq.\ \eqref{delta_e:def} with
$1/\delta_e=0.3703(1)$.}
\end{figure}

\subsection{Particle-hole symmetry: $U=-2\epsilon_d$}

For the particle-hole-symmetric case $\epsilon_d=-U/2$, we have
$\langle n_d \rangle = 1$ and $p_0=p_2=(1-f)/2$. As a result, Eq.\
\eqref{S_e:Ander:h=0} reduces to
\begin{equation}
S_e(\hloc=0) = S_2(f) + \ln 2,
\end{equation}
which increases monotonically from $\ln 2$ to $\ln 4$ as $|f-\half|$
decreases from $\half$ to $0$. Equation \eqref{S_e:slope} tells us that for
a given value of $f$, $S_e(\hloc=0)$ is greater for this symmetric
case than for any value $\epsilon_d\ne -U/2$ that leads to $p_0\ne p_2$.

In the conventional situation of on-site Coulomb \textit{repulsion}
(i.e., $U>0$), there is a monotonic evolution of the local-moment fraction
from $f=1$ at $\Gamma=0$ to $f\to\half$ for $\Gamma\gg U$, reflecting the
increased admixture of the $n_d=0$, $2$ excited configurations into the
$n_d=1$ ground state of the isolated impurity. This behavior, which is
exemplified in Fig.\ \ref{fig:Ander:symm}(a) for $r=0.4$ and two cases,
$U/D=0.005$ and $0.5$, leads to a monotonic increase in $S_e(\hloc=0)$ from
$\ln 2$ at $\Gamma=0$ to $\ln 4$ for $\Gamma\gg U$.
Both $f$ and $S_e(\hloc=0)$ increase in a smooth, featureless fashion
as $\Gamma$ rises through $\Gamma_c$.
[The mapping $U\to-U$, $\epsilon_d\to-\epsilon_d$, $\Gamma\to\Gamma$
takes one to an Anderson model with on-site \textit{attraction}, which for a
pseudogap density of states features quantum phase transitions between
local-charge and charge-Kondo phases \cite{Cheng.13}. Since the mapping
transforms $f\to 1-f$, it preserves $S_e(\hloc=0)$. We will not consider
cases $U<0$ any further in this paper.]

We have already argued that the effect of a local magnetic field $\hloc=0^+$ is
to reduce the entanglement entropy by an amount that varies near the QCP
according to Eq.\ \eqref{dS_e:Ander}. Since the QCP always occurs at a local
moment fraction $f>\half$, $S_e(\hloc=0)$ has a positive slope at
$\Gamma=\Gamma_c$. It is therefore the case that the
degeneracy-lifted entanglement entropy has a positive slope on both sides
of the quantum phase transition. In other words, it is impossible for $S_e^+$
to exhibit a peak at the QCP. In cases where the order-parameter exponent
satisfies $\beta<1$, we expect a divergence of $dS_e^+/d\Gamma$ on approach to
the QCP from the local-moment side. By contrast, for $2\beta > 1$, the
spontaneous-symmetry-breaking part of the entanglement entropy should vanish
faster than the the linear variation $S_e(\hloc=0)$ and $S_e^+$ should
therefore be essentially featureless on passing through the quantum phase
transition.

Figure \ref{fig:Ander:symm}(b) plots $S_e^+$ over a broad range $0\le
\Gamma/\Gamma_c\le 2$ for $r=0.4$, a case \cite{Ingersent.02,beta:note}
where $\beta=0.91440(2)>\half$. The curves for different values of $U$
are quantitatively different, but share the same principal features: a
smooth rise of $S_e^+$ from zero in the decoupled-impurity limit
$\Gamma/U\to 0$, with a linear variation through $\Gamma=\Gamma_c$
[not shown in detail in Figure \ref{fig:Ander:symm}(b)] leading to a
saturation $S_e^+\to\ln 4$ in the uncorrelated limit $\Gamma/U\to\infty$.

We find that for a given value of $\Delta$, $S_e^+$
in the particle-hole-symmetric Anderson model generically exceeds 
$S_e^+$ in the counterpart $\Simp=\half$ Kondo model with
the same band exponent $r$. This is a natural
consequence of the Anderson impurity being entangled with both the spin and
charge degrees of freedom of its environment.
In the limit $U/D\to\infty$, however, the form of $S_e^+(\Delta)$ for the
Anderson model converges to the corresponding function for the Kondo model.
[The Kondo function $S_e^+(\Delta)$ for several values of $r$ can be extracted
from Fig.\ \ref{fig:S_e:spin-half}(a) by rescaling the horizontal axis from
$\rho_0 J$ to $\Delta=(J-J_c)/J_c$.]
Conversely, as $U/D$ approaches zero, the critical hybridization
vanishes \cite{Logan.00} as $\Gamma_c/D\propto(U/D)^{1-r}$ while $S_e^+$
becomes ever closer to $\ln 4$ throughout the region $\Gamma\ge\Gamma_c$; a
trend that smoothly merges into the physics of a noninteracting resonant level
($U=\epsilon_d=0$) where $S_e^+=\ln 4$ for any $\Gamma>0$.

Examination of the spontaneous-symmetry-breaking part of the entanglement
entropy allows quantitative testing of Eq.\ \eqref{dS_e:Ander}. The inset to
Fig.\ \ref{fig:Ander:symm}(b) illustrates log-log plots of $\dSe(\Delta)$
that for $U/D=0.5$ and $0.005$ can be fitted to
Eq.\ \eqref{beta_e:def} with $\beta_e=1.8288(1)$, matching within estimated
error the scaling prediction of Eq.\ \eqref{beta_e:scaling} using the
aforementioned value $\beta=0.91440(2)$. Table \ref{tab:coeff-b} shows that
the ratio $\dSe/\Mloc^2$ is captured to high accuracy by
Eq.\ \eqref{dS_e:Ander}.

\begin{table}[t]
\begin{tabular}{l@{\extracolsep{2em}}l@{\extracolsep{2em}}%
                l@{\extracolsep{2em}}l}
\multicolumn{1}{c}{$U/D$} &
  \multicolumn{1}{c}{$f_c$} &
    \multicolumn{1}{c}{$2/f_c$} &
      \multicolumn{1}{c}{$b$}
\\[.5ex] \hline \\[-2ex]
0.005 & 0.637220 & 3.13863 & 3.1386(1) \\
0.05  & 0.646976 & 3.09131 & 3.0913(2) \\
0.5   & 0.689800 & 2.89939 & 2.8994(1)
\end{tabular}
\caption{\label{tab:coeff-b}
Properties at the quantum critical point of the particle-hole-symmetric
pseudogap Anderson model for $r=0.4$ and three values of $U/D$: local-moment
fraction $f_c$ (from NRG), and the predicted and actual values of
$\dSe/\Mloc^2$ in a field $\hloc=0^+$, i.e., $2/f_c$ based on Eq.\
\eqref{dS_e:Ander} and the ratio $b$ computed directly from NRG values of
$\dSe$ and $\Mloc$.}
\end{table}

We conclude our discussion of the particle-hole symmetric Anderson model by
presenting results for the effect of a finite local magnetic field.
Figure \ref{fig:Ander:symm}(c) plots the local-field-dependent part of the
entanglement entropy $\dSe(\Delta,\hloc)$ as a function of $\hloc$ for fixed
$\Delta=0$ (i.e., at the critical hybridization width). The linear variation
of the data for three different values of $U/D$ fulfills the prediction of
Eq.\ \eqref{delta_e:def}, while the fitting of quadruple-precision results
spanning fields down to $\hloc = 10^{-25} D$ (data not shown) yields an
exponent $1/\delta_e=0.37032(2)$ in excellent agreement with the value
$0.37032(4)$ deduced from Eq.\ \eqref{delta_e:scaling} using the previously
known value $1/\delta=0.18516(2)$.

\subsection{Maximal particle-hole asymmetry: $U=\infty$}

For the case $U=\infty$ of maximal particle-hole asymmetry, the parameters
entering Eq.\ \eqref{S_e:Ander:h=0} reduce to $p_2=0$,
$f=\langle n_d\rangle$, and $p_0=1-f$, so that
\begin{equation}
\label{S_e:U-inf}
S_e(\hloc=0) = S_2(f) + f \ln 2.
\end{equation}
For a given $f$, this value is smaller by $(1-f)\ln 2$ than its counterpart for
the symmetric model; indeed, Eq.\ \eqref{S_e:slope} indicates that for fixed
$f$, $S_e(\hloc=0)$ takes its smallest value when $U=\infty$ (and also when
$U=-\infty$, leading to $p_0=0$). Under variation of the local-moment fraction,
$S_e(\hloc=0)$ increases from $0$ at $f=0$ to a maximum value of $\ln 3$ at
$f=2/3$, and then decreases to reach $\ln 2$ at $f=1$.
 
For given values of $r$ and $\epsilon_d$, a quantum phase transition between
local-moment and Kondo phases will occur at some $\Gamma=\Gamma_c$. Unlike
the situation at particle-hole symmetry, where $S_e(\hloc=0)$ varies linearly
with $\Delta$ near the QCP, the variation for any $\epsilon_d\ne-U/2$ is
described by
\begin{equation}
\label{S_e-vs-Delta:h=0}
S_e(\Delta,0) - S_e(0,0)
  \simeq A |\Delta|^{1-\tilde{\gamma}} \, \mathrm{sgn} \, \Delta ,
\end{equation}
where $\tilde{\gamma}$ is the charge-susceptibility exponent at the QCP
and $A$ may be positive or negative. For $U=\infty$, this variation can be
deduced by writing
\begin{equation}
\label{chain-rule}
d S_e(\Delta,0)/d\Delta = [d S_e(\hloc=0)/df] \cdot (df/d\Delta)
\end{equation}
and noting that while $d S_e(\hloc=0)/df$ is regular near the QCP,
$df/d\Delta = \partial\langle n_d\rangle/\partial\Delta\propto
|\Delta|^{-\tilde{\gamma}}$ can be nonanalytic \cite{Pixley.12,Chowdhury.15}.
For $0.55\lesssim r < 1$, the charge susceptibility exponent $\tilde{\gamma}$ is
positive \cite{Chowdhury.15}, signaling a critical divergence of impurity charge
fluctuations on approach to the QCP from either phase. The asymmetric QCPs for
$r^*\simeq 3/8\le r\lesssim 0.55$ instead show a non-divergent charge response,
i.e., $\tilde{\gamma}=0$, a behavior also displayed at the symmetric QCPs that
exist for $\epsilon_d=-U/2$ and $0<r<\half$.

As a result of the combination of the local-moment contribution from
Eq.\ \eqref{dS_e:Ander} and the charge contribution from
Eq.\ \eqref{S_e-vs-Delta:h=0}, the degeneracy-lifted entanglement entropy will
exhibit a cusp peak at the QCP provided that (1) $dS_e(\hloc=0)/d\Delta<0$ at
$\Delta=0$ to ensure that $dS_e^+/d\Delta<0$ for $\Delta=0^+$, and
(2) $2\beta<1-\tilde{\gamma}$ to ensure that $dS_e^+/d\Delta>0$ for
$\Delta=0^-$. Previous NRG calculations have shown that condition (2) is
satisfied at the particle-hole-asymmetric QCPs for $0.42 \lesssim r < 1$
\cite{Ingersent.unpub}.
In what follows, therefore, we focus on whether condition (1) is satisfied.

\begin{figure}[t]
\includegraphics[width=2.2in,angle=-90]{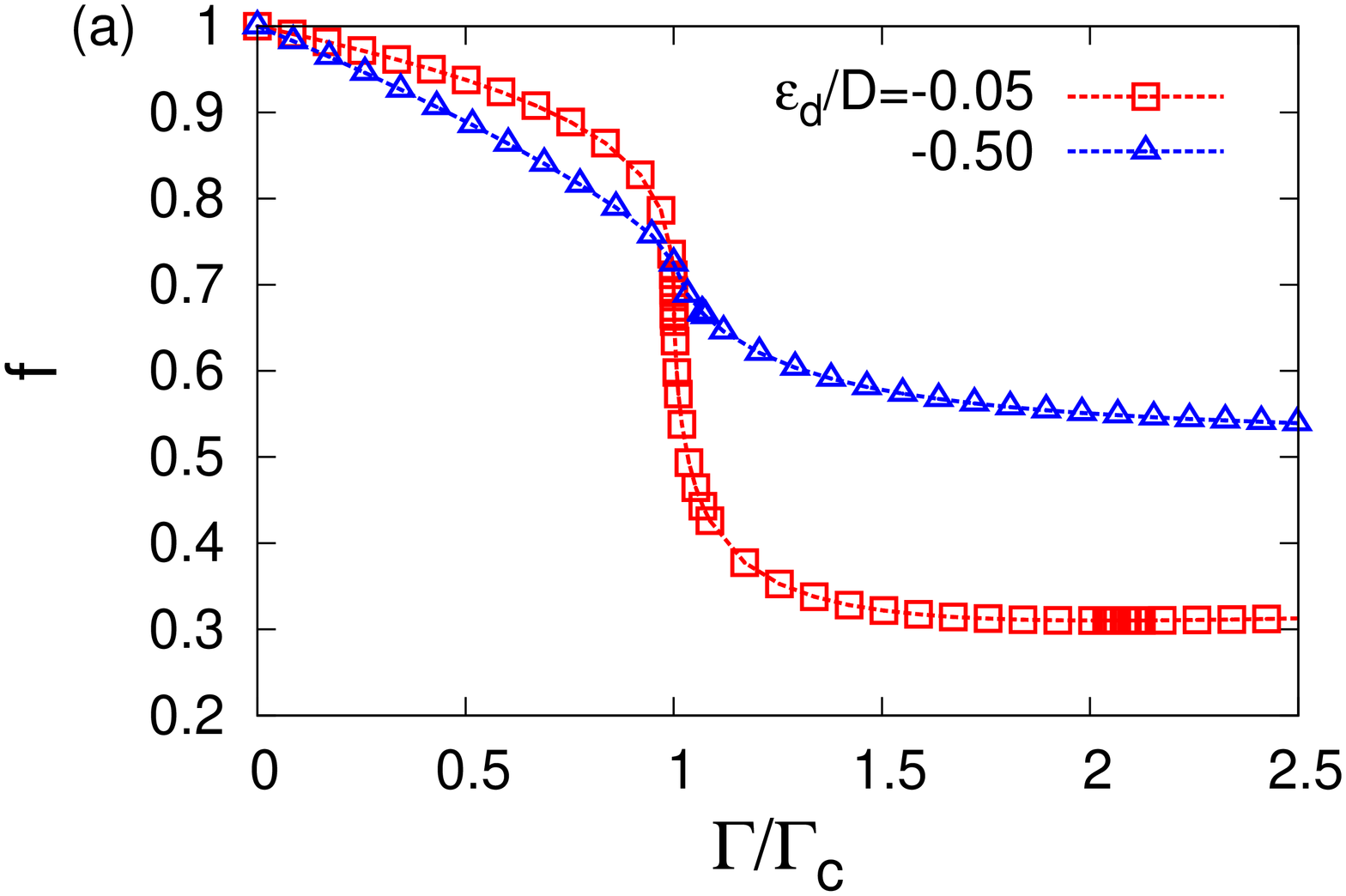} \\[-3ex]
\includegraphics[width=2.2in,angle=-90]{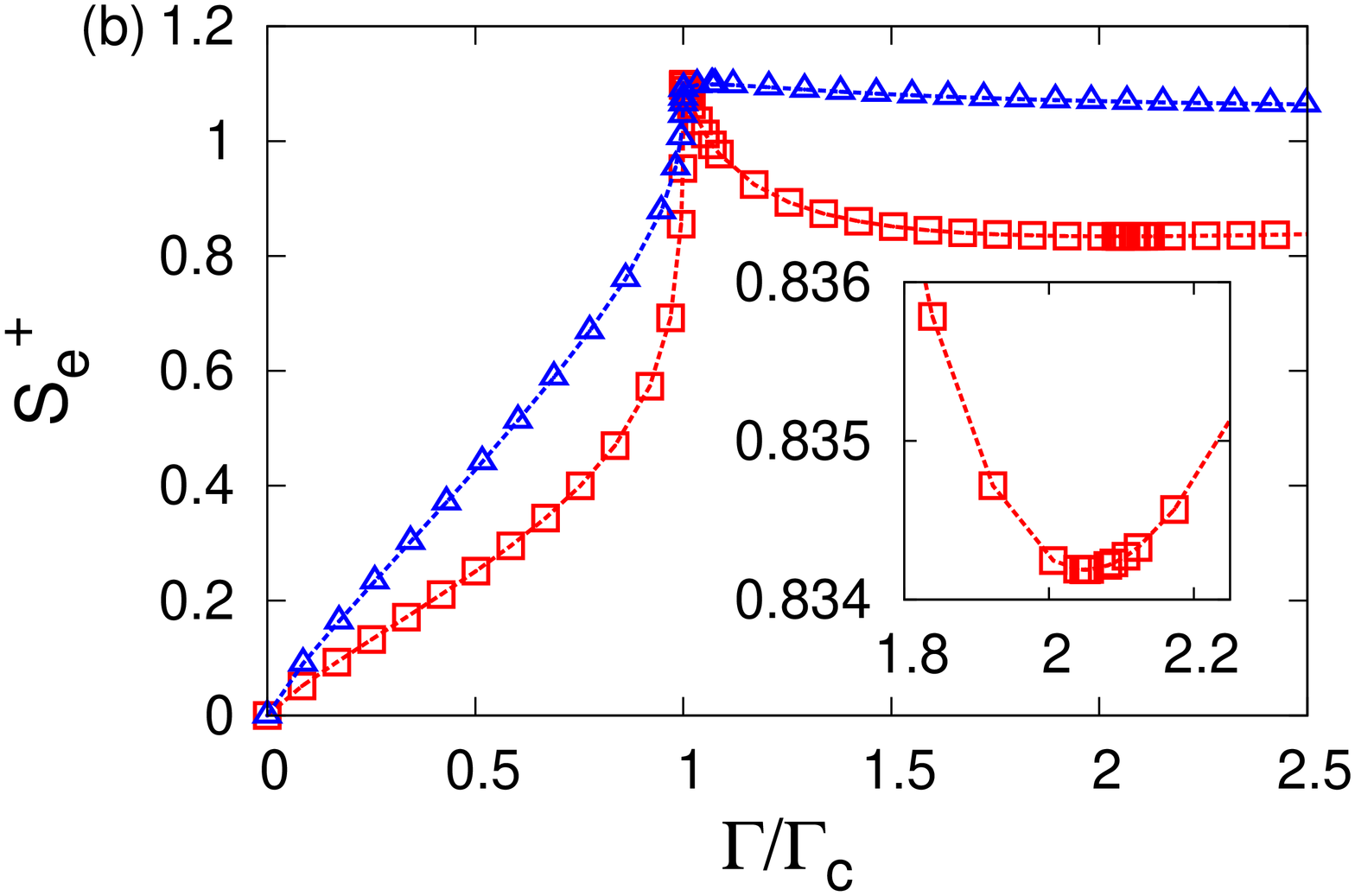}
\caption{\label{fig:Ander:U-inf}(Color online)
$U=\infty$ pseudogap Anderson model with band exponent $r=0.6$ for
$-\epsilon_d/D = 0.05$ and $0.5$:
(a) Local-moment fraction $f$ and (b) degeneracy-lifted entanglement
entropy $S_e^+$, both vs hybridization width $\Gamma$ scaled by its critical
value $\Gamma_c$. For $\epsilon_d=-0.5D$, $f$ decreases monotonically with
increasing $\Gamma$, and passes through $2/3$ at $\Gamma/\Gamma_c\simeq 1.07$,
where $S_e^+$ peaks at $\ln 3$. For $\epsilon_d=-0.05D$, $f$ drops through
$2/3$ (and $S_e^+$ peaks) at $\Gamma/\Gamma_c\simeq 1.00085$, and $f$ (and $S_e^+$) reaches
a minimum at $\Gamma/\Gamma_c\simeq 2.05$ (Inset). On the wide scale of
$\Gamma/\Gamma_c$ shown in this figure, the location of the peak in $S_e^+$
is virtually indistinguishable from the position of the quantum critical
point.}
\end{figure}

The sign of $dS_e(\hloc=0)/d\Delta$ at $\Delta=0$ can be determined using
Eq.\ \eqref{chain-rule}. From Eq.\ \eqref{S_e:U-inf} we see that
$dS_e(\hloc=0)/df$ is positive for $f<2/3$ and negative for $f>2/3$. In the
regime $\epsilon_d<0$ that admits interesting many-body physics, the
local-moment fraction has limits $f\to 1$ for $\Gamma\to 0$ and $f\to\half$
for $\Gamma\to\infty$ \cite{large-Gamma:note}. It might therefore appear
plausible that $df/d\Gamma<0$ and hence $df/d\Delta<0$ for all intermediate
values of $\Gamma$. However, NRG calculations show this assumption to be
correct only for large values of $|\epsilon_d|/D$. For smaller
$|\epsilon_d|/D$, $f$ instead has a minimum at a finite value of $\Gamma$,
beyond which it increases to approach $\half$ from below.

We present data here solely for the representative case $r=0.6$, but have
obtained qualitatively similar results for other $r$ values.
Figure \ref{fig:Ander:U-inf}(a) plots $f$ vs $\Gamma/\Gamma_c$ for
$\epsilon_d/D = -0.05$ and $-0.5$. For $\epsilon_d=-0.5 D$, $f$ decreases
monotonically with increasing hybridization width, passing through $2/3$ at
$\Gamma\simeq 1.07\Gamma_c$ where $S_e(\hloc=0)$ rises to a smooth peak at
its maximal value $\ln 3$. For $\epsilon_d=-0.05D$, $f$ drops through $2/3$
and $S_e(\hloc=0)$ peaks at $\Gamma\simeq 1.00085\Gamma_c$, barely into the
Kondo phase. However, in contrast to its behavior for $\epsilon_d=-0.5D$,
the local-moment fraction then reaches a minimum value $f\simeq 0.834\simeq
0.759\ln 3$ at $\Gamma\simeq 2.05\Gamma_c$ before rising back toward $f=\half$.
Upon further decrease of $|\epsilon_d|/D$ (not shown), the peak in
$S_e(\hloc=0)$ moves ever closer to $\Gamma=\Gamma_c$ and the minimum value of
$f$ and a related minimum in $S_e(\hloc)=0$ become deeper while the location of
these minima remains significantly above $\Gamma_c$. However, for all
$\epsilon_d < 0$ it appears that $dS_e(\hloc=0)/df>0$ and
$df/d\Delta>0$ at $\Delta=0$, meaning that the conditions are never met for the
occurrence of a peak in $S_e^+$ precisely at the QCP.

\subsection{General case}

Finally we turn to cases of intermediate particle-hole symmetry, for which
no simplification of Eq.\ \eqref{S_e:Ander:h=0} is possible. We focus once
again on the case $r=0.6$ representative of the range of band exponents
in which $2\beta<1-\tilde{\gamma}$, potentially allowing for the occurrence
of a peak in $S_e^+$ at the QCP.

\begin{figure}[t!]
\includegraphics[width=2.2in,angle=-90]{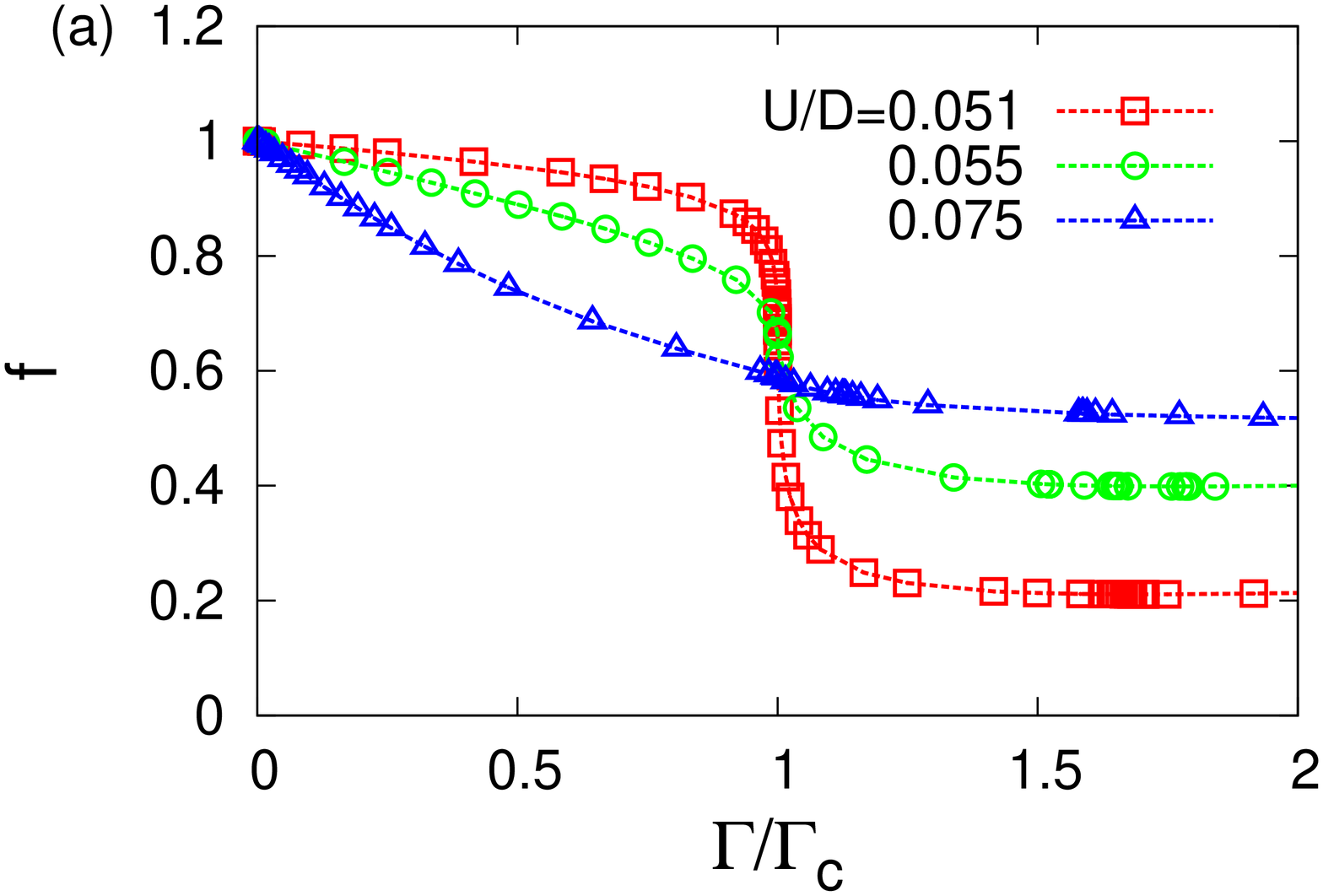} \\[-3ex]
\includegraphics[width=2.2in,angle=-90]{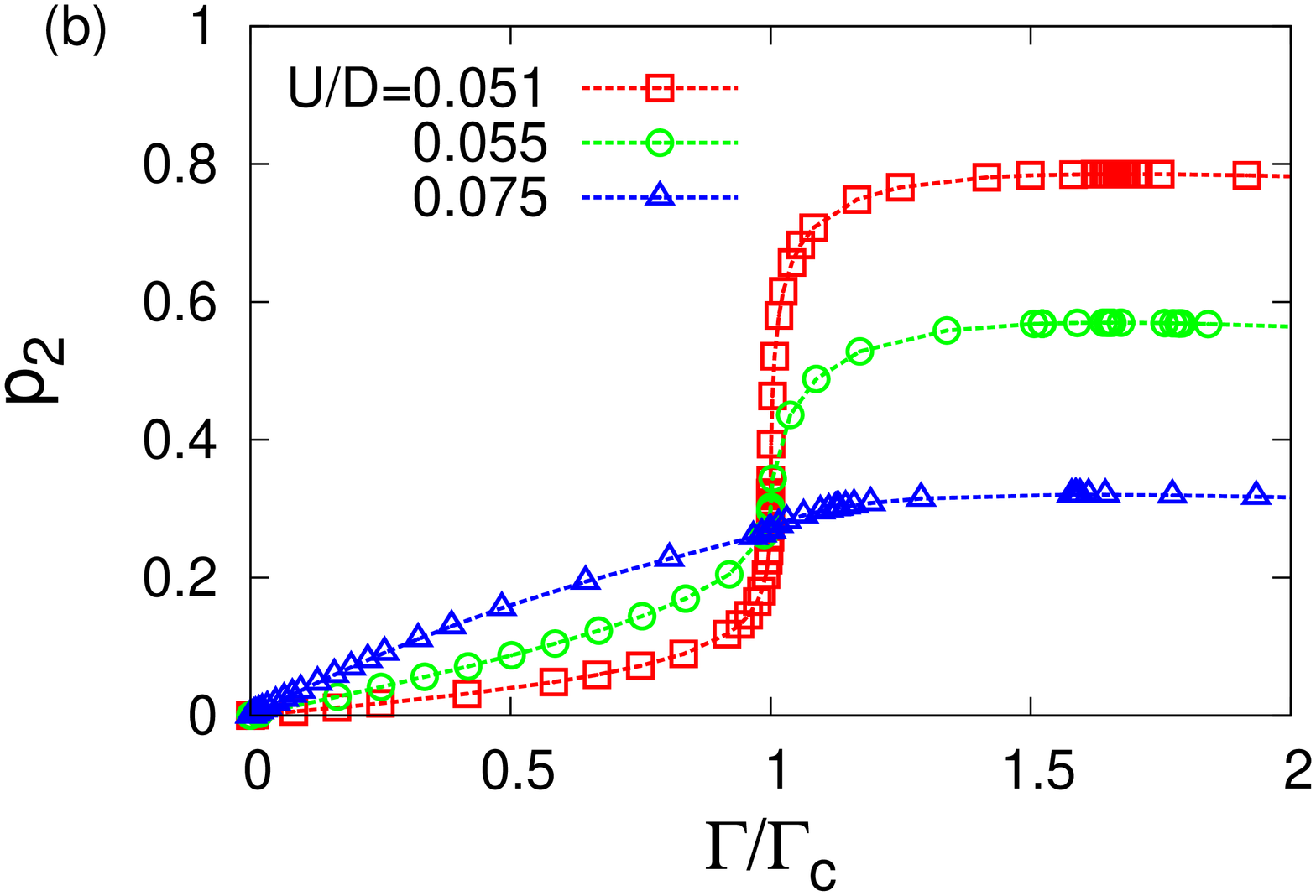} \\[-3ex]
\hspace{1ex}\includegraphics[width=2.15in,angle=-90]{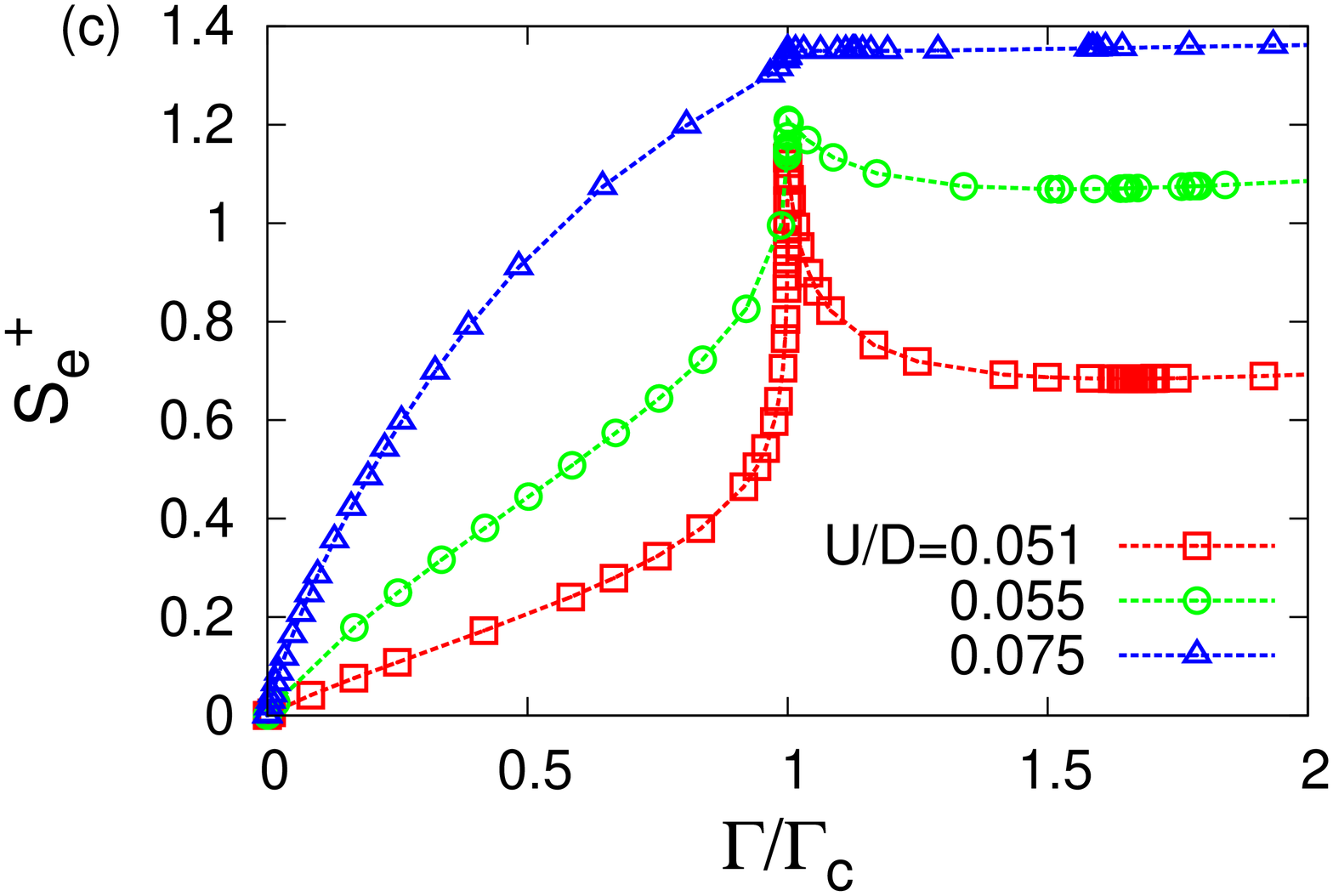}
\caption{\label{fig:Ander:gen1}(Color online)
Pseudogap Anderson model with band exponent $r=0.6$, impurity level
energy $\epsilon_d=-0.05D$, and three values of $U\ne -2\epsilon_d$:
(a) Local-moment fraction $f$, (b) double occupancy $p_2$, and (c)
degeneracy-lifted entanglement entropy $S_e^+$ vs
hybridization width $\Gamma$ scaled by its critical value $\Gamma_c$.
A peak in $S_e^+$ occurs precisely at the the quantum critical point
for $U/D=0.075$ and $0.055$, but shifted slightly into the Kondo phase
($\Gamma>\Gamma_c$) for $U=0.051D$. These cases are indistinguishable
on the wide scale of $\Gamma/\Gamma_c$ shown in this figure.}
\end{figure}

Figure \ref{fig:Ander:gen1} shows results
over a wide range of $\Gamma/\Gamma_c$ values for $\epsilon_d=-0.05D$ and
for $U/D=0.051$, $0.055$, and $0.075$. As $U$ gets closer to
$-\epsilon_d$, particle-hole asymmetry grows stronger, as do changes near
$\Gamma=\Gamma_c$ in the local-moment fraction [Fig.\
\ref{fig:Ander:gen1}(a)], the double occupancy [Fig.\
\ref{fig:Ander:gen1}(b)], and the degeneracy-lifted entanglement entropy
[Fig.\ \ref{fig:Ander:gen1}(c)]. For $U/D=0.075$ and $0.055$, $S_e(\hloc=0)$
peaks just inside the local-moment phase at $\Delta\simeq -1.2\times 10^{-3}$
and $-3.2\times 10^{-4}$, respectively. The negative slope of $S_e(\hloc=0)$
at $\Gamma=\Gamma_c$, combined with the order-parameter exponent satisfying
$2\beta<1-\tilde{\gamma}$ is sufficient to create a cusp peak in $S_e^+$ at
the QCP. By contrast, for $U/D=0.051$, $S_e(\hloc=0)$ peaks just inside the
Kondo phase at $\Delta\simeq 5\times 10^{-5}$, thereby preventing the
occurrence of any peak in $S_e^+$ at $\Delta=0$ \cite{Lambda-note}. However,
the tiny displacement of the maximum in $S_e(\hloc=0)$ from $\Delta=0$ means
that in practice it will prove very hard to distinguish cases where there
really is a peak in the degeneracy-lifted entanglement entropy at the QCP from
ones where a peak lies close by inside the Kondo phase.

Deeper into the Kondo phase, $S_e^+\equiv S_e(\hloc=0)$ exhibits a broad
minimum, which is centered at $\Gamma/\Gamma_c\simeq 1.65$, $1.52$, and
$1.13$ for $U/D=0.051$, $0.055$, and $0.075$, respectively. The minimum in
$S_e^+$ weakens as $U$ increases toward $-2\epsilon_d$, consistent with the
monotonic rise in entanglement entropy throughout the Kondo phase that is seen
at particle-hole symmetry [Fig.\ \ref{fig:Ander:symm}(c)].

\begin{figure}[t]
\includegraphics[width=2.2in,angle=-90]{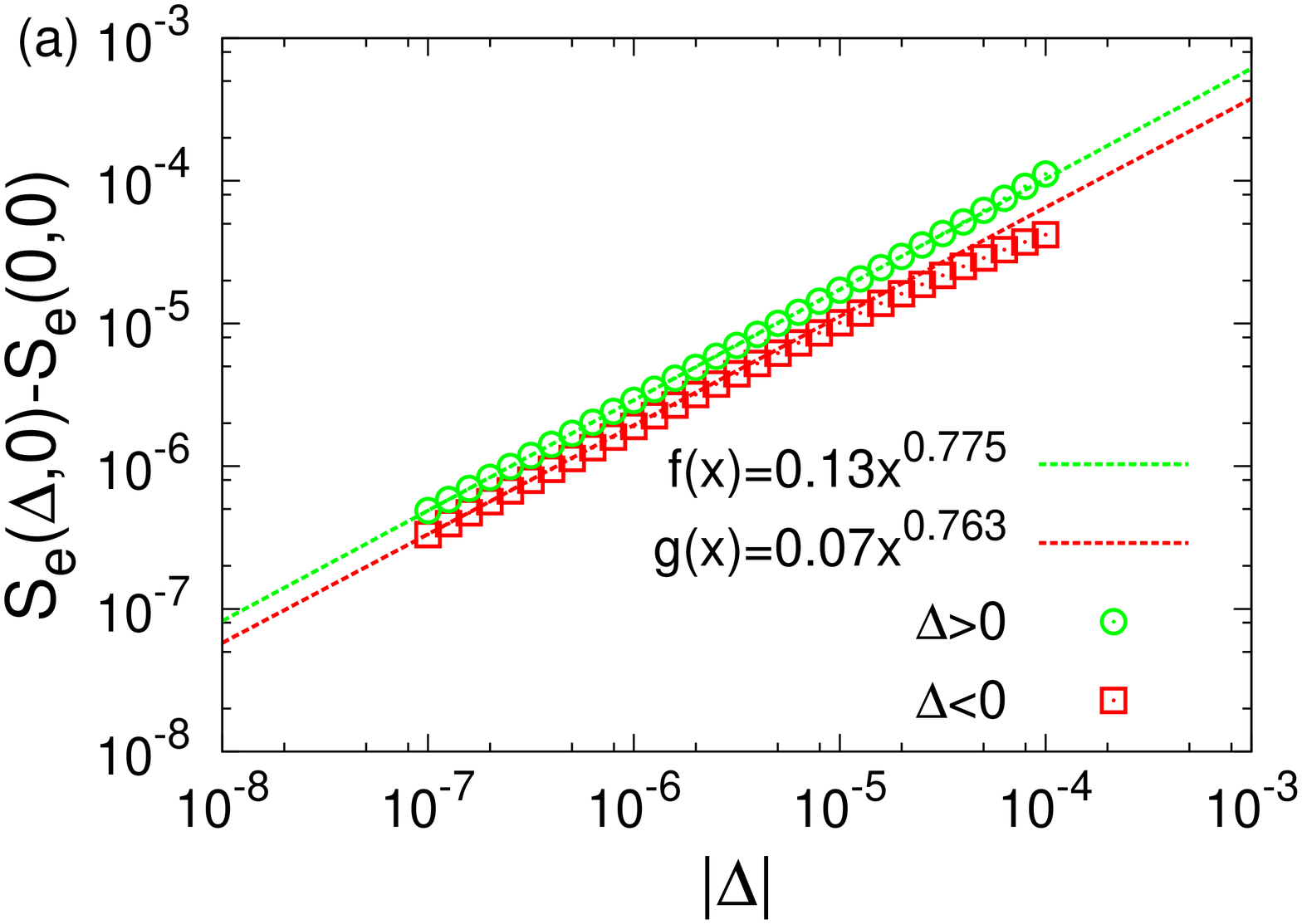}
\includegraphics[width=2.2in,angle=-90]{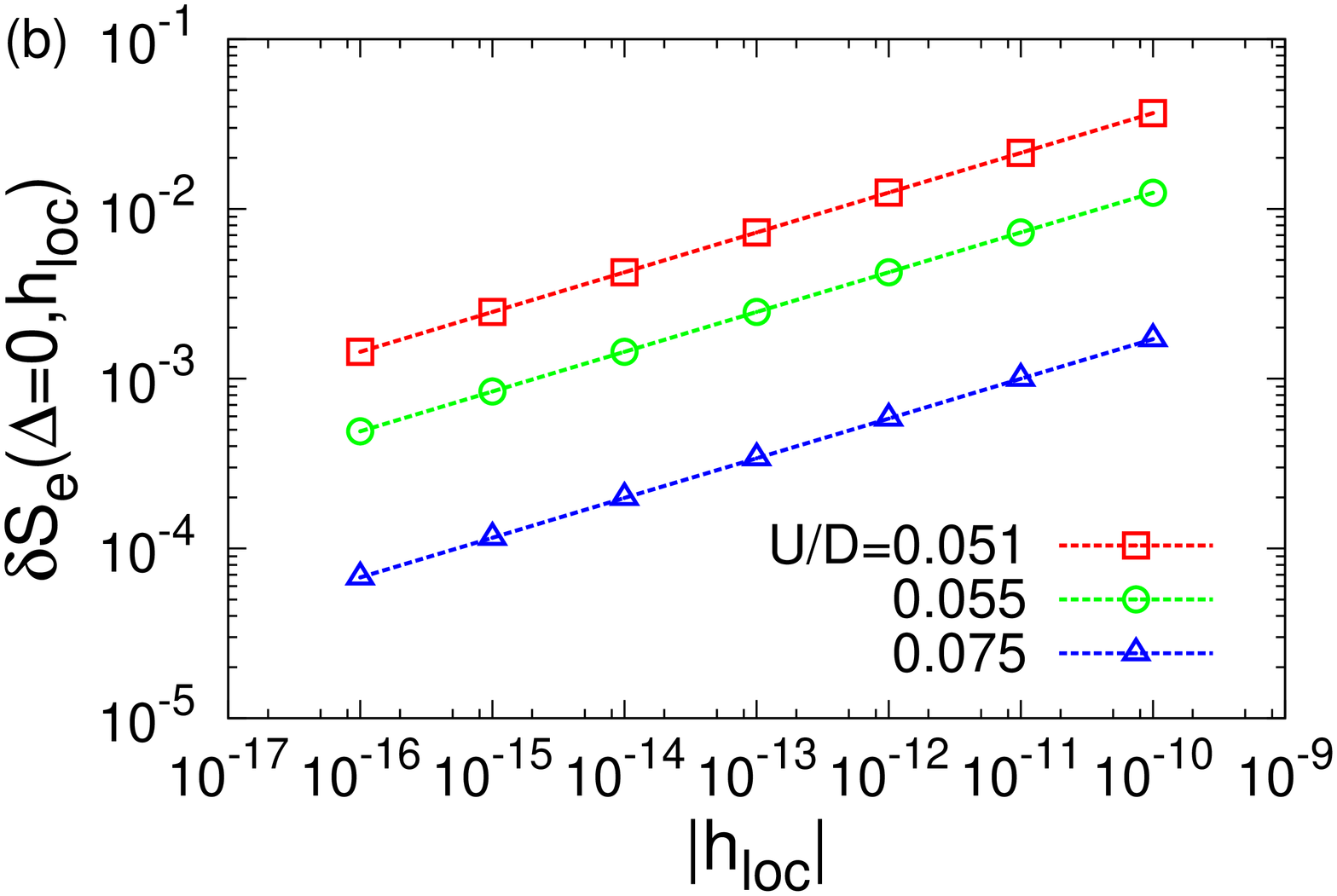}
\caption{\label{fig:Ander:gen2} (Color online)
Critical behavior of the entanglement entropy $S_e$ for the
pseudogap Anderson model with $r=0.6$, $\epsilon_d=-0.05D$, and
$U\ne -2\epsilon_d$:
(a) Shift in the zero-field entanglement entropy for $U=0.055D$ on moving
away from the quantum critical point into the local-moment phase ($\Delta<0$)
and the Kondo phase ($\Delta$). Power-law fits to Eq.\ \eqref{S_e-vs-Delta:h=0}
yield $\tilde{\gamma}=0.23(1)$.
(b) Local-field-dependent part of the entanglement entropy $\dSe$ vs $\hloc$ at
the critical hybridization width, for $U/D=0.051$, $0.055$, and $0.075$.
The data are consistent with Eq.\ \eqref{delta_e:def} with
$1/\delta_e=0.2340(1)$.}
\end{figure}

The physics near the QCP is shown in more detail in Fig.\ \ref{fig:Ander:gen2},
once again for a representative case $r=0.6$ and $\epsilon_d=-0.05D$.
Figure \ref{fig:Ander:gen2}(a) plots the critical part of the entanglement
entropy in the absence of a local magnetic field for $U=0.055D$.
The divergent charge susceptibility produces a nonanalytic variation of
$S_e(\Delta,0)-S_e(0,0)$ that is well-captured by Eq.\ \eqref{S_e-vs-Delta:h=0},
even though the power-law variation in the local-moment phase is confined to
a rather narrow region of $|\Delta|$ values due to the aforementioned peak in
$S_e(\hloc=0)$ at $\Delta=-3.2\times 10^{-4}$. The inferred exponent
$\tilde{\gamma}=0.23(2)$ is fully consistent with the
directly-computed \cite{Pixley.12,Chowdhury.15}
charge-susceptibility exponent $\tilde{\gamma}=0.210(2)$.
Figure \ref{fig:Ander:gen2}(b) shows the local-magnetic-field response at the
QCP ($\Delta=0$) for $U/D=0.051$, $0.055$, and $0.075$. Fitting to Eq.\
\eqref{delta_e:def} yields an exponent $1/\delta_e=0.2340(1)$, in excellent
agreement with Eq.\ \eqref{delta_e:scaling} given that
$2/\delta = 0.23392(8)$ based on Table \ref{tab:spin-half}.
 
\section{Discussion}
\label{sec:discuss}

One universal feature of our results is the presence of a nonzero entanglement
on entry to the local-moment (Kondo-destroyed) phase. Such a residual
entanglement implies that the ground state is not a simple product
of an impurity state and an environmental state. This result has significant
implications for theoretical and numerical descriptions of the Kondo-destroyed
phase. For example, within a large-$N$ mean field theory of the pseudogap Kondo
model \cite{Withoff.90}, the local moment is represented with fermionic spinons
$f_{\sigma}$ and the effective Hamiltonian is a resonant-level model with a
hybridization $b=\langle \hat{b}\rangle_{MF}=\langle f^{\dag}_{\sigma}
c_{0\sigma}^{\pdag} \rangle_{MF}$ (where $\hat{b}$ is a bosonic operator). At
this level, Kondo destruction corresponds to $b\rightarrow 0$, implying that
the local moment is completely free and no longer entangled with the conduction
band. Thus, such a static mean-field theory cannot reproduce the nonzero
entanglement entropy that we find in the Kondo-destroyed phase. Our results can
be understood, however, in terms of a bosonic operator $\hat{b}(\omega)$ that
has a vanishing static component and give rises to a dynamical Kondo effect. 
Our results also imply that the Kondo-destroyed phase cannot be captured in
variational quantum Monte Carlo studies of the Kondo lattice that treat $b$ as
a static variational parameter. It will be interesting to try and consider more
general variational wave functions that can treat the Kondo-destroyed phase more
accurately. 

Kondo-destroyed quantum critical points have been invoked to understand the
unconventional quantum criticality observed in experiments on heavy-fermion
metals \cite{Gegenwart.08}. As a result of the failure of the
Hertz-Millis-Moriya theory \cite{Hertz,Millis,Moriya} of the spin-density-wave
transition to describe the experimental data \cite{Coleman.01}. the concept of
local quantum criticality \cite{Si.01} has been used to understand the
energy-over-temperature scaling in the dynamic spin susceptibility, the presence
of an additional energy scale, and a jump in the Fermi-surface volume.
The theory of local quantum criticality is based on the extended-dynamical
mean-field theory of the Kondo lattice \cite{Si.01}, which finds that for
sufficiently strong quantum fluctuations the Kondo effect is indeed destroyed at
the antiferromagnetic QCP. The results of the present study imply that a continuous
loss of entanglement is expected at the Kondo-breakdown QCPs believed to occur
in certain heavy-fermion systems \cite{Si.14}.

Our results might be tested using engineered realizations of some of the
models, including single-electron transistors attached to ferromagnetic
leads \cite{Kirchner.05} and certain double-quantum-dot
systems \cite{Silva.06,Silva.08}. Equation \eqref{S_e:Ander:h=0} shows that in
the Kondo phase, knowledge of $\langle n_d\rangle$ and $\langle n_{d\uparrow}
n_{d\downarrow}\rangle$ is sufficient to determine $S_e$. Since the charge on
a quantum dot can be measured using a variety of quantum point-contact
techniques \cite{transport-book}, and it is now becoming possible to probe
double occupancy using optical spectroscopy techniques \cite{Kondo-quench},
there is a very exciting possibility of being able to extract the impurity
entanglement entropy directly from experiments.

The remarkable simplicity of the variation of the impurity entanglement
entropy near a Kondo-destruction QCP, as embodied in Eq.\ \eqref{dS_e-vs-M_loc},
can be attributed to the application of Eq.\ \eqref{S_e:def} with subsystem $A$
containing just the impurity degrees of freedom and subsystem $B$ encompassing
just the host degrees of freedom. If one considered a different partition of the
system in which host degrees of freedom were split between subsystems A and B,
then $S_e$ would probe entanglement within the host, which should be much more
sensitive than impurity-host entanglement to details such as the number of
conduction bands and/or the presence of bosonic baths. The critical behavior
near impurity QCPs of entanglement entropy defined using alternative system
partitions forms an interesting open question for future work.

\section{Conclusions}
\label{sec:conclusions}

We have studied the quantum mechanical entanglement between a magnetic
impurity and its environment in several models that feature critical
destruction of the Kondo effect. In the Kondo-destroyed phase of each model
studied, we have identified a term in the entanglement entropy varying with a
critical exponent $\beta_e =2\beta$, where $\beta$ is the critical exponent
governing the order parameter characterizing the quantum phase transition. In
addition, we have established that the response of $S_e$ to a local magnetic
field gives rise to a part of the entanglement entropy that varies with a
critical exponent $1/\delta_e=2/\delta$, where $\delta$ is the critical exponent
governing the response of the order parameter at the QCP to a local magnetic
field. We have established very generally that in Kondo models, the ratio of
the critical part of the entanglement entropy to the square of the order
parameter depends only on the magnitude of the impurity spin, and not on the
number of conduction channels or the presence of an additional bosonic bath.
In nondegenerate Anderson models,
this ratio is enhanced over its value in the $S\imp=\half$ Kondo model by a
factor of the inverse of the impurity's local-moment fraction.

Our investigation has shown the absence of any universal behavior on approach
to a Kondo-destruction quantum critical point from the Kondo (disordered) phase.
In all variants of the Kondo model that we have considered, $S_e$ remains pinned
at its maximal value of $\ln(2\Simp+1)$ throughout the Kondo phase. However, in
Anderson models the presence of charge fluctuations introduces terms that depend
on the values of $\langle n_{d\uparrow}+n_{d\downarrow} \rangle$ and
$\langle n_{d\uparrow} n_{d\downarrow}\rangle$. We have shown that in the
pseudogap Anderson model with a band exponent $r$ on the range
$0.55\lesssim r < 1$, charge fluctuations produce a nonanalytic leading
variation of $S_e$ near the QCP with a critical exponent that depends only on
$r$. Away from particle-hole symmetry, $S_e$ may rise on approach to the QCP from
the Kondo side, producing a cusp peak in $S_e$ precisely at the quantum phase
transition. However, we also find situations in which the entanglement
entropy decreases continuously, albeit nonanalytically, on passing from the Kondo
phase to the local-moment phase.

\section{Acknowledgments}

We would like to thank Kenneth Evans, M\'{a}rton Kormos, Jeremy McMinis,
Doug Natelson, Qimiao Si, and Aditya Shashi for useful discussions. This work
was supported in part by the NSF Grant No.\ DMR-1309531, Robert A.\ Welch
Foundation Grant No.\ C-1411, East-DeMarco fellowship, JQI-NSF-PFC, and
LPS-CMTC (J.H.P.), by NSF Materials World Network Grant No.\ DMR-1107814 (T.C.,
C.W., and K.I.), and by the University of Florida REU Site in Materials Physics
under NSF Grant No.\ DMR-1156737 (M.T.M.\ and J.S.).


\begin{thebibliography}{30}
\bibitem{2-chan:note}
Figure\ 30(a) of Ref.\ \onlinecite{Buxton.98} shows a line of QCPs occurring
for $0<r<r_{\max}$. Reference \onlinecite{Schneider.11} shows that this picture
is incorrect, and that there is instead weak renormalization-group flow along
the phase boundary toward a unique particle-hole-symmetric QCP.
\bibitem{LeHur:note}
Although it does not specifically identify the spontaneous-symmetry-breaking
part of the entanglement entropy, Ref.\ \onlinecite{LeHur.07} does note that
on entry to the localized phase of the sub-Ohmic spin-boson model with
$1/2 < s < 1$, the entanglement entropy decreases according to an exponent
$\nu(1-s)$ that coincides with $2\beta$, in agreement with our Eqs.\
\eqref{beta_e:def} and \eqref{beta_e:scaling}.
\bibitem{large-Gamma:note}
The large-$\Gamma$ limit of the local-moment fraction is $f=\half$,
and not $f=2/3$ corresponding to the effective degeneracy of three impurity
configurations $|0\rangle$, $|\uparrow\rangle$, and $|\downarrow\rangle$.
This is because for $\Gamma\gg|\epsilon_d|,\,D$,
the many-body ground state contains equal admixtures of
$(c_{0\uparrow}^{\dag}|0\rangle+|\uparrow\rangle)/\sqrt{2}$ and
$(c_{0\downarrow}^{\dag}|0\rangle+|\downarrow\rangle)/\sqrt{2}$.
\end{thebibliography}
\end{document}